\pdfoutput=1
\documentclass[letterpaper,preprint,11pt,fleqn,preprintnumbers,prd,onecolumn,superscriptaddress,longbibliography]{revtex4-2}

\usepackage[bookmarks=true]{hyperref}
\usepackage{graphicx,epsfig,color}
\usepackage[english]{babel}
\usepackage{amssymb,amsfonts,amsmath}
\usepackage{verbatim}

\newcommand{\be}{\begin{equation}} \newcommand{\ee}{\end{equation}}
\newcommand{\bea}{\begin{eqnarray}} \newcommand{\eea}{\end{eqnarray}}

\newcommand{\OO}{\mathcal{O}}

\newcommand{\dd}{\text{d}}
\newcommand{\chiang}{\mathcal{S}}

\numberwithin{equation}{section}

\def\d{\mathrm{d}}

\def\tens{{\cal T}_b}
\def\pot{V}
\def\Z{{\cal Z}}
\def\W{{\cal W}}
\def\k{\kappa}
\def\charge{\rho}
\def\E{E_x}
\def\mass{M_\mt{q}}
\def\Nf{N_\mt{f}}
\def\Nc{N_\mt{c}}
\def\ls{\ell_s}
\def\alphap{\ell^2_s}

\newcommand{\eps}{\epsilon}

\newcommand{\morder}[1]{{\cal O}\left(#1 \right)}

\newcommand{\pa}{\partial}
\renewcommand{\l}{\lambda}
\newcommand{\la}{\lambda}

\newcommand{\hn}{\hat n}

\newcommand{\nt}{\tilde n}

\newcommand{\mtilde}{\tilde{m}}
\newcommand{\ctilde}{\tilde{c}}
\newcommand{\LL}{\mathcal{L}}
\newcommand{\Azs}{A_t}

\newcommand{\mt}[1]{\textrm{\scriptsize #1}}

\def\cN{{\cal{N}}}

\newcommand{\bal}{\begin{aligned}}
\newcommand{\eal}{\end{aligned}}

\begin{document}

\title{Holographic approach to transport in dense QCD matter}

\author{Carlos Hoyos}
\email{hoyoscarlos@uniovi.es}
\affiliation{Department of Physics, Universidad de Oviedo and\\ Instituto de Ciencias y Tecnolog\'{\i}as Espaciales de Asturias (ICTEA)\\
Avda.~Calvo Sotelo 18, ES-33007 Oviedo, Spain}

\author{Niko Jokela}
\email{niko.jokela@helsinki.fi}
\affiliation{Department of Physics and Helsinki Institute of Physics\\
P.O.~Box 64, FI-00014 University of Helsinki, Finland}

\author{Matti J\"arvinen}
\email{matti.jarvinen@apctp.org}
\affiliation{Asia Pacific Center for Theoretical Physics, Pohang 37673, Republic of Korea}
\affiliation{Department of Physics, Pohang University of Science and Technology, Pohang 37673, Republic of Korea}

\author{Javier~G.~Subils}
\email{javier.subils@su.se}
\affiliation{Departament de F\'isica Qu\`antica i Astrof\'isica \& Institut de Ci\`encies del Cosmos (ICC), \\ Universitat de Barcelona, Mart\'i Franqu\`es 1, ES-08028, Barcelona, Spain
\\}
\affiliation{Nordita\\
KTH Royal Institute of Technology and Stockholm University\\
Hannes Alfv\'ens v\"ag 12, SE-106 91 Stockholm, Sweden}

\author{Javier Tarr\'io}
\email{tarrio@gmail.com}
\affiliation{Department of Physics and Helsinki Institute of Physics\\
P.O.~Box 64, FI-00014 University of Helsinki, Finland}

\author{Aleksi Vuorinen}
\email{aleksi.vuorinen@helsinki.fi}
\affiliation{Department of Physics and Helsinki Institute of Physics\\
P.O.~Box 64, FI-00014 University of Helsinki, Finland}


\begin{abstract}
\noindent The transport properties of dense QCD matter play a crucial role in the physics of neutron stars and their mergers, but are notoriously difficult to study with traditional quantum field theory tools. Specializing to the case of unpaired quark matter in beta equilibrium, we approach the problem through the machinery of holography, in particular the V-QCD and D3-D7 models, and derive results for the electrical and thermal conductivities and the shear and bulk viscosities. In addition we compare the bulk to shear viscosity ratio to the speed of sound and find that it violates the so-called Buchel bound. Our results differ dramatically from earlier predictions of perturbative QCD, the root causes and implications of which we analyze in detail. 
\end{abstract}

\preprint{HIP-2021-23/TH}
\preprint{APCTP Pre2021 - 018}
\preprint{NORDITA 2021-077}


\maketitle

\newpage

{\small{\tableofcontents}}

\hrulefill
\vspace{10pt}
\newpage

\section{Introduction}

Dramatic advances in neutron star (NS) observations, including the measurement of NS masses in excess of two solar masses ($M_\odot$) \cite{Antoniadis:2013pzd,Cromartie:2019kug}, the breakthrough discovery of gravitational waves (GWs) from NS mergers \cite{TheLIGOScientific:2017qsa,Abbott:2018exr}, and improvements in simultaneous mass-radius measurements (see, {\emph{e.g.}}, \cite{Nattila:2017wtj,Miller:2021qha,Riley:2021pdl}) have recently cemented the position of neutron stars (NSs) as the primary laboratory of dense Quantum Chromodynamics (QCD) matter. Through a link between the microscopic and macroscopic worlds provided by General Relativity (GR), these observations have offered valuable constraints for the properties of matter well beyond the nuclear matter saturation density of approx.~$n_s\approx 0.16~$fm$^{-3}$. 

A particularly useful model-independent framework, designed for implementing these observational constraints, is provided by controlled extrapolations and interpolations of the equation of state (EoS) of NS matter \cite{Annala:2017llu,Margalit:2017dij,Rezzolla:2017aly,Ruiz:2017due,Bauswein:2017vtn,Radice:2017lry,Most:2018hfd,Dietrich:2020efo,Capano:2019eae,Landry:2018prl,Raithel:2018ncd,Raithel:2019ejc,Essick:2019ldf, Annala:2019puf,Annala:2021gom} from the low-density realm of nuclear theory \cite{Tews:2012fj,Drischler:2017wtt,Lynn:2015jua,Holt:2014hma} to the high-density regime of perturbative QCD \cite{Kurkela:2009gj,Gorda:2018gpy,Gorda:2021znl,Gorda:2021kme}. At the moment, state-of-the-art studies of this kind are able to constrain the EoS to a good accuracy at all densities, including the most problematic region where the system is expected to transition from nuclear matter (NM) to deconfined quark matter (QM). Inside the cores of the most massive stable NSs, the properties of dense QCD matter have in addition been shown to lie closer to the expected characteristics of nearly conformal QM than to those of low-density NM, indicating the likely existence of quark cores inside at least some NSs \cite{Annala:2019puf}.

Despite all the advances described above, what is currently still lacking is a smoking-gun signal of the presence of QM inside NSs or of its creation during NS mergers \cite{Most:2018eaw}. Such a dramatic detection of QM may in fact never come from studies of the NS-matter EoS: should the deconfinement transition be only weakly first order or even a crossover at high densities and low temperatures, the EoS is likely to undergo only smooth and modest changes near the transition. In the case of a crossover transition, no well-defined critical density would even exist, but there would instead be an extended density interval where the system transitions from the hadronic phase to a deconfined one. With this possibility in mind, alongside with efforts to constrain the EoS, one should clearly search for other observable quantities that witness more significant changes when the system enters the deconfined phases of QCD.

One very promising avenue for improving our current understanding of the high-density EoS and the deconfinement transition region at high densities is a possible future observation of a GW signal from the postmerger phase of a NS merger event. This scenario has already been extensively studied using relativistic hydrodynamic simulations, revealing interesting imprints of strong first-order phase transitions (see, {\emph{e.g.}}, \cite{Most:2018eaw} for a discussion). As argued in \cite{Alford:2017rxf}, the dense and hot non-equilibrium system created in a merger is, however, not only sensitive to the EoS of NS matter, but also to the transport properties of the system, in particular its bulk viscosity. Recalling that transport quantities often experience substantial effects when the effective degrees of freedom in the system change (see, {\emph{e.g.}}, \cite{Sasaki:2008um}), being able to infer the behavior of the most important transport coefficients in both the confined and deconfined phases of QCD would be extremely valuable. Combined with data on the postmerger waveform, such information may even provide the much-needed smoking gun evidence for the creation of QM.

A severe restriction for first-principles studies of dense QCD matter is that the phase structure of the theory is not properly understood at high densities and small-to-moderate temperatures. First, the location and order of the deconfinement transition is currently an open question due to a lack of nonperturbative ab-initio tools applicable at high densities \footnote{Recall that Chiral Effective Theory and other ab-initio nuclear theory machinery are currently limited to densities below the saturation density \cite{Tews:2012fj} while lattice QCD is plagued by the infamous Sign Problem at finite density \cite{deForcrand:2010ys}.}, and second, the precise pattern of quark pairing in the deconfined phase is only known in the asymptotic limit \cite{Son:1998uk}. Given these limitations, a sound starting point is to first consider the unpaired phase when inspecting a new class of physical quantities, and then try to generalize the results to various paired phases.

In \cite{Hoyos:2020hmq}, we presented first results for a number of important transport coefficients --- the shear and bulk viscosities and the electrical and thermal conductivities --- for strongly coupled unpaired quark matter. We described the system via the holographic V-QCD framework, which is based on adding unquenched quark flavors to the Improved Holographic QCD (IHQCD) model \cite{Jarvinen:2011qe,Alho:2013hsa,Jokela:2018ers}, and additionally via the D3-D7 probe brane model for dense QCD-like quenched matter  \cite{Kruczenski:2003be,Kobayashi:2006sb}. This work followed a series of previous holographic investigations of the bulk thermodynamic properties of dense and cold quark \cite{Hoyos:2016zke,Hoyos:2016cob,Ecker:2017fyh,Annala:2017tqz,Fadafa:2019euu,Ishii:2019gta} and nuclear matter \cite{Li:2015uea,Preis:2016fsp,Kovensky:2020xif,Jokela:2020piw}, but was the first one to study transport in the high-baryon-density regime (for the high-temperature, zero-density context, see, {\emph{e.g.}}, \cite{Gursoy:2009kk,Gursoy:2010aa,Iatrakis:2014txa}). We found dramatic qualitative differences to the earlier perturbative results of \cite{Heiselberg:1993cr} --- the only previous first-principles results for these quantities --- highlighting the need for caution when using the latter in phenomenological applications in NS physics.

In the present paper, we provide a significantly more detailed account of the calculations for the transport coefficients that were only briefly covered in~\cite{Hoyos:2020hmq}. We also extend this work in several new directions: In the V-QCD setup we, among other things, provide a detailed analysis of the low-temperature behavior of both bulk thermodynamic and transport quantities in the QM phase. The low- temperature physics is governed by a ``critical'' AdS$_2\times \mathbb{R}^3$ geometry, and we argue that the low-temperature asymptotics of the transport coefficients at small $T$ can be given in terms of a transseries. We also present results for the speed of sound at finite temperature and density, and extend the analysis of viscosities to nonzero frequencies. Furthermore, we solve the transport coefficients analytically at low temperatures in the D3-D7 model; in particular, we determine the leading contribution to the bulk viscosity from the flavor degrees of freedom. Finally, we point out that this result, as well as the results for V-QCD, violate a conjectured bound for the bulk viscosity \cite{Buchel:2007mf,Gouteraux:2011qh}. 

Our article is structured as follows. In Sec.~\ref{sec:models}, we review the V-QCD and D3-D7 models applied in our analysis, after which Section \ref{sec:transportcoef} contains a detailed account of the determination of the conductivities and viscosities. In Sec.~\ref{sec:results}, we then analyze our results, while Sec.~\ref{sec:discussion} presents a brief discussion of the implications of our findings and the future prospects of improving the description of high-density transport with holographic methods. The appendixes of the article finally contain some supplementary details of the calculations, necessary for a quantitative reproduction of our results.

\section{Holographic models \label{sec:models}}

In this section, we present our generic setup that will allow us to make connection to holographic models of QCD containing fundamental matter at nonzero density. These models include both top-down options like probe brane configurations \cite{Karch:2002sh,Kobayashi:2006sb} and bottom-up approaches such as V-QCD~\cite{Jarvinen:2011qe,Alho:2013hsa}.

The action we consider consists of two separate pieces
\be\label{eq.ActionWithTachyon}
 S_\mt{total} = S_\mt{g} + S_\mt{f} \ ,
\ee
where \footnote{In order to compare with \cite{Hoyos:2020hmq}, one should set $1/(2\kappa_5^2)=N_c^2 M_{Pl}^3$, $(\tens \Z)_{\text{here}}=(N_f/N_c)\Z_{\text{there}}$.}
\bea
S_\mt{g} & = & \frac{1}{2\kappa_5^2}\int \d^5 x \sqrt{-g} \left( R - \frac{1}{2} \partial_\rho \phi \, \partial^\rho\phi - \pot(\phi,\chi) \right)  \label{eq:ActionDefsGluon} \\
S_\mt{f} & = & - \frac{\tens}{2\kappa_5^2} \int \d^5 x\, \Z(\phi,\chi) \sqrt{-\det \left( g_{\mu\nu} + \k(\phi,\chi) \, \partial_\mu \chi \, \partial_\nu\chi + \W(\phi,\chi) \, F_{\mu\nu} \right) } \ .\label{eq:ActionDefsFlavor}
\eea
The key difference to many other holographic models of QCD is the appearance of the DBI action (\ref{eq:ActionDefsFlavor}) containing a square root. We believe that polynomial actions representing fundamental degrees of freedom would lead to results qualitatively different from those reported here. We also note that we write our action in five-dimensional language. It should be relatively straightforward to apply our techniques in higher-dimensional models if the reduction to five dimensions can be done explicitly.

Let us now discuss the different components of the actions above and their origins. In top-down setups, the $S_\mt{g}$ piece descends almost entirely from the closed string sector of type II supergravity by reducing on a five-dimensional compact manifold, and is thus related to the glue sector of a gauge theory with gauge group of rank $\Nc$. The scalar field $\phi$ is identified with the dilaton, which in holographic contexts is dual to the Yang--Mills operator and its coupling constant. The potential term $V(\phi,\chi)$ originates in the higher-dimensional Ramond--Ramond and Neveu--Schwarz fluxes and the compact geometry upon which one reduces. This reduction could in principle give rise to several scalars, $\varphi_i$, that parametrize the compact manifold. Here we are omitting these in our description for simplicity, since considering them will not give new insights in our results, and their inclusion is in principle straightforward albeit tedious. Notice, however, that the potential depends explicitly on a second scalar, $\chi$, which we will refer to as the tachyon. Its kinetic term is included in $S_\mt{f}$, as we shall discuss shortly. The dependence of the potential on this field will not have any important consequences in our results, but we include it explicitly since this dependence is present in smeared setups \cite{Nunez:2010sf}. In V-QCD, however, we will restrict to $V=V(\phi)$ configurations: The $S_\mt{g}$ sector of V-QCD is motivated by noncritical five dimensional string theory, but with a form for this potential that is chosen to match with known features of Yang-Mills theory as we will discuss below.

The second part of the action, given by $S_\mt{f}$, corresponds to the open string sector determined by the Dirac--Born--Infeld (DBI) action of a set of source branes in a top-down picture, and models the flavor sector of the theory. The scalar field $\chi$ would then describe how the flavor branes, whose tension we have written as $\tens/(2\kappa_5^2)$, are embedded in the higher-dimensional geometry. In bottom-up approaches, such as V-QCD, this scalar is referred to as the tachyon, since $S_\mt{f}$ is inspired by Sen's tachyonic action \cite{Sen:2002in,Bigazzi:2005md,Bergman:2007pm,Casero:2007ae,Jokela:2009tk,Iatrakis:2010jb,Iatrakis:2010zf}, and we borrow this nomenclature. The field strength $F_{\mu\nu}=\partial_\mu A_\nu-\partial_\nu A_\mu$ in turn gives the dynamics of a U(1)$_\mt{B}$ gauge field corresponding to the conserved baryonic charge of the dual field theory. The couplings $\Z$, $\k$, and $\W$ depend on the dilaton and tachyon, and in top-down setups they are completely determined by supergravity, whereas in V-QCD they are (in analogy to the choice of the potential $V(\phi)$) chosen to reproduce specific features of QCD. From now on we will omit the arguments in the different potentials $\pot$, $\Z$, $\k$, and $\W$.

Extremizing Eq.~\eqref{eq.ActionWithTachyon}, we obtain equations of motion that are more concisely expressed in terms of the matrix
\be\label{eq.GammaMatrix}
\Gamma_{\mu\nu} \equiv g_{\mu\nu} + \k\, \partial_\mu \chi \partial_\nu\chi + \W\, F_{\mu\nu} \ ,
\ee
the inverse components of which we write as $\Gamma^{\mu\nu}$; notice that this does not equal $(g^{-1}\Gamma g^{-1})^{\mu\nu}$. These equations of motion are given in Appendix~\ref{app.eoms}.
In the following, we will consider a homogeneous, rotationally-invariant background in the presence of a non-trivial potential along the time-direction for the gauge field, which is a requisite for systems at finite density:
\be\label{eq.Ansatz}
\d s^2  =  g_{tt}(r) \d t^2 + g_{xx}(r) \d \vec x ^2 + g_{rr}(r) \d r^2 \ ,  \quad
A  = A_t(r) \d t \ , \quad 
\phi  = \phi(r) \ , \quad 
\chi  = \chi(r) \ . 
\ee

A first integration for the gauge field can be readily obtained. To do this, notice that within our Ansatz
\be
 \Gamma^{[\mu\nu]} = \frac{g_{xx}^3}{\det \Gamma} A_t' \,\left( \delta^{\mu t}\delta^{\nu r} - \delta^{\mu r}\delta^{\nu t} \right)
\ee
such that from the equation of motion for the U(1)$_\mt{B}$ gauge field in \eqref{eq.EquationsOfMotion} we can define the conserved 4-dimensional current
\be\label{eq.Current}
 J^m = \frac{\tens}{2\kappa_5^2} \, \Z \,\W \sqrt{-\Gamma} \, \Gamma^{[m r]} = (\charge,\vec 0)^m
\ee
to obtain
\be
 F_{rt} = A_t'(r) =  \frac{2\kappa_5^2 \, \charge}{\W} \frac{ \sqrt{-g_{tt}} \sqrt{g_{rr} + \k \,\chi'^2}}{\sqrt{\left( 2\kappa_5^2 \, \charge\right)^2 + \tens^2 \Z^2 \W^2 g_{xx}^3}} \ .
\ee
From the holographic dictionary, the constant of integration $\charge$ corresponds to the conserved U(1)$_\mt{B}$ charge density, whereas the chemical potential is given by the value of the vector potential at the boundary $\mu = A_t(\infty)$.

From now on we assume the presence of a nonzero temperature, such that our Ansatz \eqref{eq.Ansatz} behaves near the horizon $r_\mt{h}$ as 
\be\label{eq.HorizonBehavior}
\begin{aligned}
A_t^\mt{h} & \simeq {\cal O}(r-r_\mt{h}) \ , & g_{tt}^\mt{h} \simeq -\alpha (4\pi T) (r-r_\mt{h}) +{\cal O}(r-r_\mt{h})^2 \ , \\
g_{rr}^\mt{h} & \simeq \frac{\alpha}{4\pi T} (r-r_\mt{h})^{-1} +{\cal O}(r-r_\mt{h})^0 \ , & g_{xx}^\mt{h}  \simeq (2\kappa_5^2 s/4\pi)^{2/3} + {\cal O}(r-r_\mt{h}) \ ,
\end{aligned}
\ee
where $s$ is the entropy density, $T$ the temperature, and $\alpha$ an undetermined constant that can be absorbed in a redefinition of the holographic radius $r\in[r_\mt{h},\infty)$.

We will now proceed to particularize our former expressions to specific cases. In this paper, we will focus on two different setups: the probe D3-D7 system and V-QCD.

\subsection{Probe D3-D7}\label{sec:setupD3D7}

First, let us focus on probe brane setups, in which the action $S_\mt{g}$ for the glue degrees of freedom is solved without taking into account the backreaction of $S_\mt{f}$, so that there is no dependence in the potential $\pot$ on the tachyon $\chi$. The corresponding background is fixed and given as input when solving the dynamics associated to the tachyon and the U(1)$_\mt{B}$ gauge field from $S_\mt{f}$. In the field theory dual, this is interpreted as working in a quenched approximation for flavors.

The results thus obtained can only be trusted to first order in a parameter that depends on the ratio between coupling constants, $\tens$. To include effects from the backreaction is clearly interesting \cite{Cremonini:2017qwq}, albeit also conceptually intricate due to the appearance of a Landau pole, which forces one to set up the holographic dictionary at a finite cut-off surface \cite{Bigazzi:2009bk,Bigazzi:2011it,Bigazzi:2011db,Bigazzi:2013jqa,Faedo:2016cih}.

It is far from obvious that the action for the D3-D7 intersection can be written in the 5-dimensional form presented in (\ref{eq.ActionWithTachyon}). We will hence derive it explicitly starting from ten dimensions.
For the D3-D7 intersection, the background corresponds to a solution of type IIB supergravity sourced by a Ramond--Ramond 5-form flux and a constant dilaton ($\phi=0$). The solution reduces to an AdS$_5\times$S${}^5$ metric
\be\label{eq.AdS5S5metric}
\d s_{10}^2 = G_{MN} \d X^M \d X^N = g_{tt} \d t^2 + g_{xx} \d \vec x^2 + g_{rr} \d r^2 + L^2 \left( \d \theta^2 + \cos^2 \theta\, \d \Omega_3^2 + \sin^2 \theta\, \d \phi^2 \right) \ ,
\ee
where
\be
g_{tt} = - \frac{r^2}{L^2} \left(1-\frac{r_\mt{h}^4}{r^4}\right) \ , \quad g_{xx} = \frac{r^2}{L^2} \ , \quad g_{rr}=-g_{tt}^{-1} \ , \quad r_\mt{h} = \pi T L^2 \ .
\ee
Consequently, the gluonic part of the action can be written in this case as
\be 
S_\mt{g} = \frac{1}{2\kappa_{5}^2}\, \int \, \dd^{5}x \,  \sqrt{-g}\, \left(R + \frac{12}{L^2} \right)\,.
\ee

The value of the five-dimensional Newton's constant $\kappa_{5}$ is related to the ten-dimensional one and the volume of S$^5$ as $2\kappa_5^2 = 2\kappa_{10}^2\, (\pi^3L^5)^{-1}$.
The thermodynamic quantities associated to this black brane solution are
\be\label{eq.AdS5S5thermo}
s = \frac{4\pi^4\, L^3}{2\kappa_5^2} \, T^3 \ , \qquad \varepsilon = \frac{3\pi^4\, L^3}{2\kappa_5^2} \, T^4 \ , \qquad p=-f =  \frac{\pi^4\, L^3}{2\kappa_5^2} \, T^4 \,,
\ee
for the entropy ($s$) and energy ($\varepsilon$) densities as well as for the pressure ($p$) and free energy density ($f$). 

Concerning the flavor sector, we add to this background a stack of $\Nf$ D7-branes wrapping the $\textrm{S}^3\subset \textrm{S}^5$. We made this three-dimensional sphere explicit in our expression for the metric \eqref{eq.AdS5S5metric}. These D7-branes are considered extended along the four non-compact spatial directions \cite{Karch:2002sh}, and their dynamics are determined by the DBI action in terms of the pull-back of the metric to the eight-dimensional worldvolume of the D7-brane
\be \label{eq:action_DBI}
S_\mt{DBI} = - T_7 \Nf\, \int\,  \d^8\xi\, e^{-\phi}\sqrt{-\det \left( G_{MN} \frac{\d X^M}{\d \xi^m} \frac{\d X^N}{\d \xi^n} + 2\pi\alphap \, F_{mn} \right)} \ .
\ee

In the above action, $X^M$ refers to the 10-dimensional coordinates and $\xi^m$ to the worldvolume coordinates on the D7-branes. 
Also, $T_7$ is the tension of a single D7-brane, given by $T_7 = {1}/{((2\pi\ell_s)^7g_s\ell_s)}$ as a function of the string length $\ls$ and the string coupling $g_s$. The embedding is fully determined by the functional dependence of the ten-dimensional angle $\theta$ on the worldvolume coordinates. Put differently, the embedding is described by the profile of the angle $\theta$ in the worldvolume of the D7-branes, which is taken to depend solely on the radial coordinate $r$, so that $\theta=\theta(r)$.

As discussed above, in order to describe dense phases we need to introduce a U(1)$_\mt{B}$ field in the worldvolume of the D7-branes given by
\be
A = A_t(r) \, \d t \ ,
\ee
which will be dual to the chemical potential and charge density of the system. Taking all these ingredients into account, the DBI action of the D7-branes \eqref{eq:action_DBI} finally reduces to
\be\bal
S_{\mt{f}} = S_\mt{DBI} & = - \cN \, \int \d^4 x \, \d r\, r^3 \cos^3\theta \, \sqrt{1 + f\, r^2\, \theta'^2 - (2\pi\alphap)^2 F_{rt}^2} \\
& = -\cN\, V_3\, \beta\,  \, \int \, \dd r\, r^3 \cos^3\theta \, \sqrt{1 + f\, r^2\, \theta'^2 - \frac{4\pi^2\, L^4}{\lambda} F_{rt}^2} \equiv
 \int\, \mathcal{L}_{\mt{f}} \ ,
\eal\ee
where 
\be
 \cN = 2\pi^2\, \Nf\, T_7  = \frac{\Nc\, \Nf\, \lambda}{(2\pi)^4\, L^8 } \ 
\ee
and $V_3$ and $\beta = 1/T$ correspond to the volume of three dimensional space and the integral over the temporal component, respectively. In the last equality, we have made use of the holographic dictionary, which relates the supergravity parameters to gauge theory quantities in the following way:
\be \label{eq:dictionary}
\frac{L^4}{\ls^4} = \lambda \ , \quad  \lambda = 4\pi g_s \Nc \ .
\ee
From the action we have just presented, we can obtain the equations of motion that $A_t(r)$ and $\theta(r)$ have to fulfil. As usual, we can integrate readily for the electric field by defining
\be\label{eq:D3D7_electric_field_eq}
(V_3\, \beta )^{-1}\,  \frac{\delta S_\mt{f}}{\delta A_t'} \equiv \rho \quad \Rightarrow \quad A_t' = \rho\, \frac{\lambda}{2\pi L^2 } \frac{\sqrt{1+f(r) r^2 \theta'^2}}{\sqrt{\lambda \, \rho^2  +  4 \pi^2 L^4\cN^2 r^6 \cos^6\theta}} \ ,
\ee
so that $\rho$ is the charge density in the field theory side. On the other hand, we get a second order differential equation for the angle $\theta(r)$

\be\label{eq:D3D7_embedding_eq}
 \frac{\delta S_\mt{f}}{\delta \theta} = \frac{\partial \mathcal{L}_{\mt{f}} }{\partial\theta(r)} -  \partial_r \frac{\partial \mathcal{L}_{\mt{f}} }{\partial\theta'(r)}=0 \ .
\ee

Even though in some particular cases there are analytic solutions to \eqref{eq:D3D7_electric_field_eq} and \eqref{eq:D3D7_embedding_eq} which we will discuss later, in general we will need to solve the differential equations numerically. Near the boundary, the solutions for the electric potential and embedding profile read
\bea \label{eq.assymD3D7}
 \theta & = & \frac{\mtilde}{r} + \frac{\ctilde}{r^3}   + \ldots  \nonumber\\
 A_t & = & \mu  - \frac{ \lambda\, \rho}{8\pi^2 \cN L^4} \frac{1}{r^2} + \ldots \label{eq:D3D7_UV_exp}
\eea
with $\mu$ the chemical potential, $\rho$ the charge density, and $\mtilde$ and $\ctilde$ related to the mass of the quark and quark condensate via the standard holographic dictionary, respectively: 

\be\label{eq:D3D7_quark_mass_condensate}
M_\mt{q} = \frac{ \sqrt{\lambda}}{2 \pi}\,\frac{ \mtilde}{L^2} \ \ , \ \ 
\langle \overline{\psi}\psi \rangle = - \frac{\Nf\Nc\,\sqrt{\lambda}}{4\,\pi^3\, L^6}  \left( \ctilde - \frac{\mtilde^3}{6} \right) \ .
\ee
Notice that our conventions differ from those in \cite{Mateos:2007vn}, such that $\lambda_{[\text{here}]} = 2 \lambda_{[\text{there}]}$.

By matching our action \eqref{eq:ActionDefsFlavor} to the DBI one for the D7-brane, identifying the tachyon with the embedding function, 
\be
\chi=\theta(r) \ , 
\ee
we obtain (we set already $\phi=0$)
\be\label{eq.D3D7functions}
\frac{\tens}{2\kappa_5^2} = T_7 \, \Nf \, \int_{\textrm{S}^3} \omega_3 = 2 \pi^2\,  T_7 \,  \Nf = \cN\ , \qquad \W = 2 \pi \alphap \ , \qquad \k=1 \ , \qquad \Z = L^3\,\cos^3 \theta \,,
\ee
whose relation to gauge theory quantities is again given by \eqref{eq:dictionary}.

As mentioned before, in order to find the transport properties of the D3-D7 probe system, we will need to find the precise embedding realized in each case (at different temperatures, charge densities, quark masses, etc.) by solving Eqs.~\eqref{eq:D3D7_electric_field_eq} and \eqref{eq:D3D7_embedding_eq}. The way this is done follows \cite{Kobayashi:2006sb,Mateos:2007vn} and is described in Appendix~\ref{app:D3D7 computations}.

\subsection{V-QCD}\label{sec:setupVQCD}

V-QCD represents a class of bottom-up models~\cite{Jarvinen:2011qe}, constructed to reproduce a number of features of full QCD including both gluon and flavor dynamics with full backreaction. The structure of the model is therefore richer than those of most other bottom-up models studied in the literature. The model is based on one hand on improved holographic QCD (IHQCD)~\cite{Gursoy:2007cb,Gursoy:2007er}, inspired by 5D noncritical string theory, for the gluon sector, and on a mechanism for adding brane actions with tachyon condensate~\cite{Bigazzi:2005md,Casero:2007ae}, inspired by a space-filling D$4-\overline{\textrm{D}4}$-brane configuration, for the flavor sector. As the model is loosely based on a brane configuration, its action resembles that of the D3-D7 model in many ways. That is, in the absence of CP-odd and flavor-dependent sources the action of V-QCD can be cast in the form of Eqs.~\eqref{eq.ActionWithTachyon} and~\eqref{eq:ActionDefsGluon}-\eqref{eq:ActionDefsFlavor}. 

The glue part of the action,~Eq.~\eqref{eq:ActionDefsGluon}, equals the action for IHQCD~\cite{Gursoy:2007cb,Gursoy:2007er} if the potential is chosen appropriately. As already pointed out above, the dilaton field $\phi$ is dual to the $\mathrm{Tr}F^2$ operator. We denote
\be
  \lambda = e^{\sqrt{3}\phi/\sqrt{8}}\ ,
\ee
since in this normalization the source term of $\lambda$ at the boundary matches the 't Hooft coupling in QCD. %
Note that the definition of $\lambda$ is therefore quite different from its counterpart in the D3-D7 model above. In particular, the coupling runs in V-QCD, i.e., its value depends on the holographic coordinate. We have, however, decided to use the same symbol for different bulk quantities in the two models, since their interpretation in field theory is the same. 

The potential $V(\lambda)$ is chosen to only depend on $\lambda$, and at the boundary we require that \footnote{Note that most literature on the IHQCD and V-QCD models uses the opposite sign for the potential, $V_g(\lambda)= -V(\lambda)$.}
\be \label{eq:VUVlim}
 V(\la) \to -\frac{12}{L^2} \,,\qquad  (\la \to 0)\ ,
\ee
with corrections suppressed by powers of $\la$. The resulting geometry is asymptotically AdS$_5$ at the boundary, and $L$ is the asymptotic AdS radius. The physically more interesting IR domain of the model is specified by setting the behavior of $V(\la)$ at intermediate and large values of $\la$. In particular choosing the asymptotics 
\be \label{eq:VIRlim}
 V(\la) \sim \la^{4/3}\sqrt{\log\la} \,,\qquad  (\la \to \infty)\ ,
\ee
this choice produces confinement and magnetic screening as well as linear radial trajectories for the glueball spectrum~\cite{Gursoy:2007cb,Gursoy:2007er}. 
See Appendix~\ref{app:vqcd} for the specific choice of the potential used in this article.

The flavor action~\eqref{eq:ActionDefsFlavor} in V-QCD is chosen following Sen~\cite{Sen:2003tm}, {\emph{i.e.}}, the potential is taken to be an exponential of the squared tachyon field $\chi^2$:
\be \label{eq.Senpotential}
 \mathcal{Z}(\lambda,\chi) = V_{{\mt{f}}0}(\lambda)e^{-\chi^2}  \ .
\ee
Recall that the tachyon field is dual to the quark bilinear $\bar qq$ in QCD. The condensation of the tachyon in the bulk, driven by the exponential potential, therefore gives rise to chiral symmetry breaking in QCD~\cite{Bigazzi:2005md,Casero:2007ae}. The corresponding source is the (flavor-independent) quark mass, which we set to zero in this article for simplicity.

Apart from the function $V_{{\mt{f}}0}(\la)$, the flavor sector contains the potentials $\mathcal{W}(\la)$ and (even though it will not be important in this article) $\kappa(\la)$.
For these potentials, we first require consistency of the solutions: that the IR singularity is indeed of the good kind \cite{Gubser:2000nd}, and that the tachyon IR flow of the chirally broken solutions destroys the flavor action in the IR, as suggested by the choice of the tachyon potential in~\eqref{eq.Senpotential}~\cite{Jarvinen:2011qe,Arean:2013tja}. This corresponds to the annihilation of the D$4-\overline{\textrm{D}4}$ pair in the deep IR, therefore realising the tachyon decay picture of Sen. 

We also require that the radial trajectories of the meson spectra are asymptotically linear~\cite{Arean:2012mq}, that the mass gaps in the meson spectra grow with the quark mass~\cite{Jarvinen:2015ofa}, and that the phase diagram of the model as a function of $T$ and $\mu$ has qualitatively reasonable behavior~\cite{Ishii:2019gta}. These comparisons demonstrate that the natural choice for potentials are power laws in $\lambda$ with exponents inspired by string theory~\cite{Arean:2013tja}, potentially modified by logarithmic corrections.

In the UV, we require, as usual, that the asymptotics of the bulk fields match with the dimensions of the dual operators, which constrains the asymptotic values of the potentials. Moreover, we require that the holographic RG flow of the coupling $\lambda$ (up to two-loop level) and the quark mass (up to one-loop level) matches with that in pQCD, which constrain the UV corrections in $\lambda$ of the potentials. The idea is that constraining the UV behavior ``by hand'' using pQCD gives the best possible boundary conditions for the IR physics, where holography can lead to nontrivial predictions.

In the end, we tune the remaining degrees of freedom in the glue sector~\cite{Gursoy:2009jd} and in the flavor sector~\cite{Jokela:2018ers}  by comparing to the lattice data for the thermodynamics of pure Yang-Mills and QCD, respectively. The potential sets we use for the numerical results presented below are those constructed in~\cite{Alho:2015zua,Jokela:2018ers,Ishii:2019gta}.  
Specifically, we will present results for three choices of these potential sets, 
given by the fits {\bf{5b}}, {\bf{7a}}, and {\bf{8b}} in~\cite{Jokela:2018ers}, respectively (see Appendix~\ref{app:vqcd}). 

For the V-QCD setup, we work in the Veneziano limit, where the parameter $\tens =\Nf/\Nc$ is kept finite (but $\Nf \to \infty$ and $\Nc \to \infty$). In this limit, there is full backreaction between the gluon ($S_\mt{g}$) and flavor ($S_\mt{f}$) sectors, which also means that --- unlike in the probe D3-D7 case --- it is difficult to obtain any analytic solutions even for the background, including the scalar fields and the geometry. The temporal component of the gauge field $A_t$ may be eliminated analytically in the same way as for D3-D7, but the other fields are found by using purely numerical methods, which make use of analytically known expansions of the background in the IR or at the horizon as boundary conditions. Observables are in turn computed using known expansions of the fields at the boundary. See~\cite{Alho:2012mh,Alho:2013hsa} for details.

We find that the V-QCD action admits four kinds of homogeneous background solutions~\cite{Alho:2012mh}. First, there are two types of geometries: horizonless ``thermal gas'' solutions where the IR geometry ends in a ``good kind'' of singularity according to Gubser's classification~\cite{Gubser:2000nd}, and black hole solutions with a horizon in the IR~\cite{Gursoy:2008bu,Gursoy:2008za}. Second, in the absence of explicit chiral symmetry breaking realized through the source of the tachyon, there are solutions with either zero tachyon or spontaneously formed tachyon ``hair'' in the bulk. All of these four types of background solutions also admit generalizations to finite charge~\cite{Alho:2013hsa} due to the gauge field in~\eqref{eq:ActionDefsFlavor}. In the setup we will study in this article, only two of these solutions appear in the phase diagram determined by dominant solutions: the tachyonic thermal gas solutions (dual to the chirally broken confined QCD phase) and tachyonless black hole solutions (dual to the chirally symmetric deconfined quark matter phase).

Apart from these homogeneous phases, the full phase diagram model may contain additional structure. Nuclear matter within a simple approximation was considered in~\cite{Ishii:2019gta,Ecker:2019xrw,Jokela:2020piw} \footnote{This requires taking into account the non-Abelian structure of the action, which is not included in Eq.~\eqref{eq:ActionDefsFlavor}.}. As expected, the nuclear matter phase appears at low temperatures and intermediate chemical potentials.
The transport properties are, however, only nontrivial in the black hole phases because nontrivial thermodynamics is $1/\Nc$ suppressed in the confining phases. Therefore we will concentrate on the (chirally symmetric) black hole phase, which is dual to unpaired quark matter in this article. Since we are working at zero quark mass, this means that $\chi=0$ everywhere in this phase. That is, the relevant V-QCD action simplifies to
\be
S_\mt{V-QCD}   =  \frac{1}{2\kappa_5^2}\int \d^5 x \sqrt{-g} \left( R - \frac{4}{3} \frac{(\partial \la)^2}{\la^2}  - \pot(\la) \right)   - \frac{\Nf}{2\Nc\kappa_5^2} \int \d^5 x\, V_{{\mt{f}}0}(\la) \sqrt{-\det \left( g_{\mu\nu}  + \W(\la) \, F_{\mu\nu} \right) } \ . 
\ee

\section{Transport coefficients}\label{sec:transportcoef}

In this section, we present the derivation of general formulas for the conductivities and viscosities in the class of models we study in this work. This requires only knowledge of the background geometry at the black hole horizon, which greatly simplifies the calculation. 

\subsection{Conductivities}\label{sec:conductivities}

We are interested in the DC transport coefficients obtained as the response of the charge and heat currents to homogeneous electric fields and temperature gradients
\be
J^i=\sigma^{ij}E_j-\alpha^{ij}\nabla_j T \ , \qquad  Q^i=-\kappa^{ij}\nabla_j T+T\alpha^{ij}E_j \ .
\ee
The set of coefficients $\sigma$, $\alpha$, and $\kappa$ correspond to the electric, thermoelectric, and heat conductivities, respectively. As we are considering homogeneous and isotropic states, all the conductivity tensors are proportional to $\delta^{ij}$. In principle, all the transport coefficients can be determined through Kubo formulas involving correlators of the current and energy-momentum tensor. Although the calculation is straightforward using holography, there is an even simpler approach that involves evaluating the solutions at the black brane horizon and can be used even in non-homogeneous states \cite{Donos:2014cya}.
Our derivation will follow closely this work, with some deviations that we will highlight below.

We start by considering a fluctuation with linear-in-time sources 
\be\label{eq.FluctuationAnsatz}
\begin{aligned}
\delta g_{\mu\nu} & = \left[\,  g_{tt}\, \zeta_x\, t + h_{tx}(r) \right] \left( \delta_{\mu t}\delta_{\nu x} + \delta_{\mu x}\delta_{\nu t} \right) + h_{rx}(r)  \left( \delta_{\mu r}\delta_{\nu x} + \delta_{\mu x}\delta_{\nu r} \right)  \\
\delta A_\mu & = \left[ - \left(\E - A_t\, \zeta_x \right) t  + a_x(r) \right] \delta_{\mu x} \ .
\end{aligned}
\ee
We identify $\zeta_x=-\nabla_x T/T$ with the gradient of temperature and $\E$ with the electric field. We have aligned both of these in the $x$-direction to be able to cancel the electric and heat currents against each other, such that a translation-invariant, static configuration arises, with no change of momentum.

The equations of motion at linear order in the fluctuations are given in the Appendix, see Eq.~\eqref{eq.EquationForFluctuations}. These equations are two Einstein equations (an equation for $h_{tx}$ and a constraint coming from the radial Einstein equation) and one equation for the gauge field $a_x$, with no time dependence once the background is set on-shell.

The constraint gives a radially conserved quantity, which when evaluated at the horizon with the horizon behavior \eqref{eq.HorizonBehavior} gives rise to the relation
\be\label{eq.Seebeck}
\charge \,\E + s\,T\, \zeta_x = 0 \quad \Rightarrow \quad \E = \frac{s}{\charge} \nabla_x T \ .
\ee
In the original derivation of \cite{Donos:2014cya}, an axion field that breaks translational invariance in the $x$-direction is considered, and the presence of this field suffices to avoid the appearance of a relation like \eqref{eq.Seebeck}. If the translations are only broken spontaneously, the relationship \eqref{eq.Seebeck} emerges, as shown in \cite{Gouteraux:2018wfe}. This condition corresponds to a zero net force from the combined temperature gradient and electric field, as we will argue below; a derivation of this result in relativistic hydrodynamics can be found in Appendix C of \cite{Hoyos:2020hmq}.

Besides the radial Einstein equation that gave rise to \eqref{eq.Seebeck} there are other two radially conserved quantities. The first one is the current in the $x$-direction as read from Eq.~\eqref{eq.Current} and the second one corresponds to the heat current \cite{Donos:2014cya}
\be\label{eq.ElectricandHeatCurrents}
\begin{aligned}
 J^x & = \charge \frac{h_{tx}}{g_{xx}} + \frac{\W}{2\kappa_5^2} \frac{\sqrt{-g_{tt}}}{g_{xx}}  \frac{\sqrt{(2 \kappa_5^2\, \charge)^2 + \left( g_{xx} \right)^3 \tens^2\, \W^2 \Z^2}}{\sqrt{g_{rr} + \k \,\chi'^2}} a_x' \\
Q^x & =  \frac{\left(-g_{tt}\right)^{3/2} \sqrt{g_{xx}}}{2\kappa_5^2 \sqrt{g_{rr}}} \partial_r \left( \frac{h_{tx}}{g_{tt}} \right) - A_t J^x   \ .
\end{aligned}
\ee
The equations of motion can be solved near the horizon by
\be\label{eq.FluctuationSolutionsHorizon}
a_{x} = \frac{\E - A_t(r)\, \zeta_x}{4\pi T} \log(r-r_\mt{h}) + \ldots \ , \qquad h_{tx} = - \frac{\zeta_x\, g_{tt}(r)}{4\pi T} \log (r-r_\mt{h}) + \frac{2 \kappa^2 H^x}{4\pi T\, (g_{xx}^\mt{h})^{1/2}} + \ldots \ ,
\ee
with $H^x$ an undetermined constant of integration. Evaluating at the horizon by plugging \eqref{eq.HorizonBehavior} and \eqref{eq.FluctuationSolutionsHorizon} into these expressions, we get
\be\label{eq.ElectricCurrent}
\begin{aligned}
 J^x & = \frac{\charge \, H^x}{s\, T}+ \sigma \, \E  \\
Q^x & = H^x \ ,
\end{aligned}
\ee
where
\begin{equation}\label{eq:sigma}
    \sigma=\frac{\W_\mt{h}}{2\kappa_5^2} \frac{ \sqrt{(2 \kappa_5^2\, \charge)^2 + \left( g_{xx}^\mt{h} \right)^3 \tens^2\, \W_\mt{h}^2 \Z_\mt{h}^2}}{ g_{xx}^\mt{h} } \ .
\end{equation}

Near the boundary, we impose that no non-normalizable mode is turned on to ensure that the explicit sources given in \eqref{eq.FluctuationAnsatz} are the only ones. In this case, the expansion of the $tx$ component of the metric at the boundary reads
\be
\delta h_{tx} = \frac{r^2}{L^2}\left(0+ \frac{L\kappa_5^2}{2}(H^x+\mu J^x)\frac{L^4}{r^4}+\ldots\right).
\ee
The coefficient of this term determines the momentum density in the $x$-direction, $T^{0 x}=H^x+\mu J^x$, which is consistent with the identification of $H^x$ as the heat current \cite{Donos:2014cya}. If there was a net force, the momentum density would increase linearly with time, but the Ansatz is consistent with a time-independent momentum density. This implies that the net force acting on the plasma by the electric field and the temperature gradient is zero, as we argued above. 

For convenience, we will in the following work in the rest frame of the fluid, where the momentum density vanishes. Then, we have the condition
\be
Q^x=-\mu J^x \ .
\ee
Taking this into account, we can write
\be
 J^x - \frac{\charge}{s\,T}Q^x = \frac{\varepsilon+p}{s\,T}J^x = \sigma \, \E  \ ,
\ee
which, in turn, gives rise to the DC conductivity
\be\label{eq.DCconductivity}
  \sigma^{xx} = \frac{s\,T}{\varepsilon+p} \sigma \ .
\ee

In relativistic hydrodynamics, there is a single transport coefficient that determines both the electrical and thermal conductivities, which fixes the relation \cite{Hoyos:2020hmq}
\be\label{eq.kappa_equation}
\kappa^{xx}=\frac{\mu s}{\rho}\sigma^{xx} \ .
\ee
Note that these transport coefficients have a slightly different interpretation than the one we usually assign them. The charge conductivity is normally defined in the absence of a temperature gradient and the heat conductivity in the absence of an electric field (while the thermal conductivity is defined in the absence of a charge current). However, in a clean system with unbroken translation invariance, the usual DC transport coefficients would be formally divergent because the forces produced by any of the sources will introduce an acceleration of the elements of the fluid that would make the currents grow without bound. We have avoided this issue by tuning the electric field and temperature gradient in such a way that the induced forces compensate each other and the total acceleration of the fluid vanishes. In this case the induced currents remain finite, but the definition of the transport coefficients has been modified.

\subsection{Viscosities}\label{sec:viscosities}

Next, we explain our prescription for determining the shear and bulk viscosities within the models introduced in Sec.~\ref{eq.ActionWithTachyon}. Concretely, we do this by evaluating the appropriate quantities at the horizon $r=r_\mt{h}$ as specified below. 

The shear viscosity can be computed from the entropy density following the usual Kovtun-Son-Starinets (KSS) relation \cite{Kovtun:2003wp}, which is valid for any matter coupled to Einstein gravity  \cite{Iqbal:2008by}
\be
\eta = \frac{s}{4\pi} = \frac{(g_{xx}^\mt{h})^{3/2}}{2\kappa_5^2}+\frac{s_\mt{f}}{4\pi} \ .
\ee
In the V-QCD model $s_\mt{f}=0$, as the effect of the flavors is already included in the background geometry. In the D3-D7 model, the flavor contribution $s_\mt{f}$ is specified in \eqref{eq:sf}.

The bulk viscosity is on the other hand computed following the method developed by Eling and Oz \cite{Eling:2011ms}. We expect the Eling-Oz derivation to be valid in the D3-D7 model, since flavors are quenched similarly to some examples studied in \cite{Eling:2011ms}. In the V-QCD model, we have checked that the bulk viscosity computed in this way coincides with the one obtained through the conventional method of extracting it from perturbations of the black hole \cite{Gubser:2008sz,Buchel:2011wx}. 
Below, we present only the Eling-Oz derivation since it is considerably simpler. The alternative derivation for the V-QCD model is detailed in Appendix~\ref{app:finiteom}, as we also use it to find the frequency-dependence of the viscosities.

The computational framework is based on the fluid/gravity correspondence \cite{Bhattacharyya:2008jc}, where the fields in the background geometry are given a dependence on the spacetime coordinates along the field theory directions and a solution is systematically constructed expanding in derivatives. Fortunately, in order to obtain the bulk viscosity we do not need to find the explicit form of the derivative corrections, as they will start affecting transport coefficients only at second order.

The hydrodynamic equations of the fluid in the field theory dual are obtained by projecting the Einstein equations \eqref{eq.EquationsOfMotion}  on the black hole horizon. The projection is performed with the null vector $\ell^\mu$ transverse to the horizon
\be\label{eq.NullVectorProjection1}
R_{\mu\nu} \ell^\mu \ell^\nu - \frac{1}{2} \left( \ell^\rho \partial_\rho \phi  \right)^2 + \frac{1}{2} \tens \Z \frac{\sqrt{-\Gamma}}{\sqrt{-g}} \ell_\mu \ell_\nu \Gamma^{\mu\nu} = 0 \ .
\ee
The first two terms are the same as the ones found in the original derivation \cite{Eling:2011ms}. The last term can on the other hand be cast in a similar form by specializing to the Ansatz \eqref{eq.Ansatz}, for which the following equality holds: 
\be\label{eq.TachyonGradientProjection}
\Gamma^{\mu\nu} \ell_\mu \ell_\nu = - \k \, \frac{\det g}{\det \Gamma} \left( \ell^\rho \partial_\rho \chi  \right)^2 \ .
\ee
Then, the projected Einstein's equations \eqref{eq.NullVectorProjection1} become
\be\label{eq.NullVectorProjection2}
R_{\mu\nu} \ell^\mu \ell^\nu - \frac{1}{2} \left( \ell^\rho \partial_\rho \phi  \right)^2 - \frac{1}{2} \tens \Z\, \k \frac{\sqrt{-g}}{\sqrt{-\Gamma}} \left( \ell^\rho \partial_\rho \chi  \right)^2 = 0 \ .
\ee
This is now of the form used in  \cite{Eling:2011ms} for several scalar fields. It follows directly from their analysis that the bulk viscosity is determined by the horizon values
\be
\frac{\zeta}{\eta} = \left( s \frac{\partial \phi_\mt{h}(s,\rho)}{\partial s} + \rho \frac{\partial \phi_\mt{h}(s,\rho)}{\partial \rho} \right) ^2 
+ \tens \Z_\mt{h}\,\k_\mt{h} \frac{\sqrt{-g_\mt{h}}}{\sqrt{-\Gamma_\mt{h}}}
\left( s \frac{\partial \chi_\mt{h}(s,\rho)}{\partial s} + \rho \frac{\partial \chi_\mt{h}(s,\rho)}{\partial \rho} \right) ^2
\ ,
\ee
which, after using \eqref{eq.DCconductivity}, gives
\be\label{eq.BulkViscosityFinal}
\frac{\zeta}{\eta} \quad = \quad  \left( s \frac{\partial \phi_\mt{h}(s,\rho)}{\partial s} + \rho \frac{\partial \phi_\mt{h}(s,\rho)}{\partial \rho} \right) ^2  +  \frac{2\kappa_5^2}{\left( g_{xx}^\mt{h} \right)^{1/2}} \, \frac{\k_\mt{h}}{\W_\mt{h}^2} \, \sigma \,  \left( s \frac{\partial \chi_\mt{h}(s,\rho)}{\partial s} + \rho \frac{\partial \chi_\mt{h}(s,\rho)}{\partial \rho} \right) ^2 \ .
\ee

The above expression is the most general one for the type of models we are studying. However, in the V-QCD model we are working in the chirally symmetric phase of the theory featuring massless quark flavors, so the tachyon field vanishes $\chi=0$ and only the first term in \eqref{eq.BulkViscosityFinal} contributes to the bulk viscosity. On the other hand, in the D3-D7 model the dilaton vanishes $\phi=0$ and it is the second term that produces a non-zero bulk viscosity. In this second case we work at leading order in $\Nf/\Nc$, so in practice we do not include the entropy density of flavors in \eqref{eq.BulkViscosityFinal} and $s$ only contains the glue contribution determined by the area of the black hole horizon.

\section{Results}\label{sec:results}

Now that the expressions for the DC transport coefficients have been written down, we can determine their values in the models discussed earlier. We begin this by first discussing the D3-D7 system and turn to the V-QCD setup afterwards.

\subsection{Probe D3-D7}\label{sec:resultsD3D7}

The expressions for the different transport coefficients we are interested in for the D3-D7 system follow straightforwardly from the relations we found in the previous section. Firstly, we can plug the relations between the top-down parameters and the functions in the five-dimensional model into Eq.~\eqref{eq:sigma}. Doing so, we obtain
\be\label{eq.ProbeConductivity}
 \sigma =  \sqrt{\left( \frac{2 \charge}{\sqrt{\lambda}\,\pi\, T^2}\right)^2 + \frac{\Nf^2 \Nc^2 T^2}{16 \pi^2}    \cos^6 \theta_\mt{h}} 
\ee
which coincides with the result in \cite{Karch:2007pd} by sending the electric field there, $e$, to zero and identifying $\charge$ with $D$ in that reference. Note that this conductivity is of order $\Nf\, \Nc$, which corresponds to a calculation in the probe approximation. This can be made more explicit by extracting several factors from the charge density 
\be
\charge = \sqrt{\lambda} \frac{\Nf \Nc}{8\pi^3} \bar \charge 
\ee
in such a way that $\bar \charge$ is independent of $\Nf$, $\Nc$, or $\lambda$, whereby \eqref{eq.ProbeConductivity} becomes
\be
\sigma =  \frac{\Nf \Nc T}{4\pi} \sqrt{  \left(\frac{\bar \charge}{\pi^3\, T^3}\right)^2 + \cos^6 \theta_\mt{h} } \ .
\ee

As we saw in the previous Section, this is not yet the electrical conductivity since an additional prefactor appears in the final result \eqref{eq.DCconductivity}, namely 
\be
\sigma^{xx} = \frac{sT}{\varepsilon+p}\, \sigma \ .
\ee
At finite temperature and to leading order in $\Nf/\Nc$ ({\emph{i.e.}}, the background contribution), this prefactor equals $1$. However, at zero temperature the leading order vanishes and the prefactor is non-trivial. Consequently, we will keep the next-to-leading order in $\Nf/\Nc$ in the prefactor in our plots, given that we expect the contribution from flavors to overcome the glue contribution at small enough temperatures. Note that $\Nf/\Nc$ corrections to $\sigma$ play differently, as the leading part never vanishes and corrections can be taken into account only after computing the backreaction of the brane on the background.

Once the electrical conductivity is known, the thermal conductivity follows directly from (\ref{eq.kappa_equation}). The results for thermal and electrical conductivity are subsequently shown in Fig.~\ref{fig:ConductivitiesD3D7}. 

Once the conductivities have been found, we can move on to the computation of the viscosities. 
Concerning the shear viscosity, we already mentioned it can be directly obtained from the relation $\eta/s = 1/(4\pi)$ in this case. Therefore, the contribution to the shear viscosity from the flavor sector,
\be
\eta_\mt{f} = \frac{s_\mt{f}}{4\pi}\ ,
\ee
can be directly obtained from its contribution to the entropy density $s_\mt{f}$, whose computation in this system is reviewed in Appendix~\ref{app:D3D7 computations}.

For the bulk viscosity \eqref{eq.BulkViscosityFinal}, we get a vanishing contribution from the constant dilaton.
Using $\eta/s=1/(4\pi)$ and \eqref{eq.AdS5S5thermo}, we then obtain from \eqref{eq.BulkViscosityFinal}
\be\label{eq.BulkViscosityProbe}
\zeta =  \frac{\lambda\,  \sigma \, T^2}{4}  \, \left( s \frac{\partial \theta_\mt{h}(s,\rho)}{\partial s} + \rho \frac{\partial \theta_\mt{h}(s,\rho)}{\partial \rho} \right) ^2 
\ee
for the probe calculation. Note that only the gluonic contribution has been considered in $\eta$, so that the terms $(\Nf/\Nc)^2$ are properly neglected.

The evaluation of the squared term above can be performed numerically in general, as discussed in Appendix~\ref{app:D3D7 computations}, in addition to which analytical expressions can be obtained in some particular limits. The result for the bulk and shear viscosities is shown in Fig.~\ref{fig:ViscositiesD3D7}. From now on we will restrict to the case $T\ll \mass$, which is relevant for the study of neutron stars.

\subsubsection{Probe D3-D7: Leading behavior at low temperatures}

The interesting limit of zero temperature is recovered as $r_\mt{h}$ approaches to zero. Following \cite{Karch:2007br}, we are able to provide analytical expressions for many of the quantities we have been discussing.

First, we perform a Legendre transform which simplifíes the computations. Recall that the grand canonical potential is proportional to the DBI action
\begin{equation}
\frac{S_{\text{\tiny DBI}}}{V_3}\ =\ \beta \int_{r_\mt{h}}^\infty \dd r\, \LL (\theta,\dot{\theta},A_t,\dot{A}_t;r_\mt{h}) = - \beta\,  \Omega \ ,
\end{equation}
where $V_3 = \int_{\mathbf{R}^3}\dd^3x$ is the volume of three-space. Consequently, the free energy density is given by
\be
f = \Omega + \rho\, \mu =  \int_{r_\mt{h}}^\infty \dd r\, \left(-\, \LL (\theta,{\theta'},A_t,A'_t;r_\mt{h}) + \rho A'_t(r) \right) \equiv \int_{r_\mt{h}}^\infty \dd r\, \tilde{\LL} (\theta,{\theta'},A_t,A'_t;r_\mt{h}) \ ,
\ee
where we use the identity $\mu=A_t(\infty)-A_t(r_\mt{h})=\int_{r_\mt{h}}^\infty\dd r\, A_t'$. From this, the entropy density can be computed as 
\begin{equation}\label{eq:entropylowT}
 s\   =\  -\frac{\partial f(T,\rho)}{\partial T}  \ =\  - \pi L^2 \frac{\partial f(r_\mt{h},\rho)}{\partial r_\mt{h}}
 = \ \pi L^2 \tilde{\LL}\, \Big|_{r=r_\mt{h}} -
\pi L^2
\int_{r_h}^\infty \dd r\, \left(\frac{\dd}{\dd r_\mt{h}}\right)_{\rho,m}\left[ \tilde{\LL} (\theta,{\theta'},A_t,A'_t;r_\mt{h}) \right]\ ,
\end{equation}
where the derivative in the last term is defined through a variation that keeps both $\rho$ and the quark mass (which is the only source here) fixed. 

Because the gluonic contribution to the entropy density vanishes at $T=0$ (see Eq.~\eqref{eq.AdS5S5thermo}), the expression in~\eqref{eq:entropylowT} corresponds to the flavor contribution to entropy in this limit. Furthermore, the second piece of this expression vanishes, as we show in Appendix~\ref{app:D3D7 computations}. Consequently,
\begin{equation}
s_{\text{\tiny{f}}}\, (T=0) \, =  \,   \pi L^2 \, \tilde{\LL}\, \Big|_{r\, =\, 0} \,,
\end{equation}
so we only need to find an expression for $\tilde\LL$ evaluated at the limiting solution for which $r_\mt{h}=0$. In order to find such embedding, it is quite convenient to perform the following change of coordinates in our background metric \cite{Karch:2007br}:
\be\label{eq.changeofvariables}
y = r \sin\theta \ , \qquad z= r \cos \theta \ .
\ee
In these coordinates, the function $y(z)$ will describe the embedding of the D7-branes, whose action reduces to
\be\label{eq:D3D7ZT_action}
S_\mt{DBI} = - \cN \int \d^4x\, \d z\,  z^3 \sqrt{1 + (\partial_z y)^2 - \frac{4\pi^2}{\lambda} (\partial_z A_t)^2}\,.
\ee
Then we have
\begin{equation}
\tilde{\tilde{\LL}}(y,\dot y, A_t,\dot A_t) = -z^3 \cN \sqrt{1 + \dot y^2 -(2\pi\alpha')^2 \dot A_t^2} + \rho \, \dot A_t\,,
\end{equation}
where dots stand for differentiation with respect to $z$ and we are putting two tildes on $\LL$ to account for integrating over $z$ rather that $r$. Note that the integrand in Eq.~\eqref{eq:D3D7ZT_action} does not depend explicitly on $A_t(z)$ (as Eq.~\eqref{eq:D3D7_electric_field_eq}) or on $y(z)$. This implies that there are two conserved quantities
\be
L^{-4}\, \frac{\partial \tilde{\tilde{\LL}}}{\partial \partial_z y} = -C \,, \qquad  L^{-4}\, \,  \frac{\partial \tilde{\tilde{\LL}}}{\partial \partial_z A_t} = \rho\,,
\ee
the second of which is nothing but the charge density.
From these conserved quantities we get the corresponding equations of motion
\be\bal
\partial_z A_t & = \frac{\lambda\, \rho}{2\pi L^2 \sqrt{ \lambda \rho^2 + 4\pi^2 L^4 (z^6\cN^2 - C^2)}} \\
\partial_z y & =  \frac{2\pi L^{2}\, C}{ \sqrt{ \lambda \rho^2 + 4\pi^2 L^4 (z^6\cN^2 - C^2)}} \ .
\eal\ee

Fortunately, the two first order differential equations above can be integrated exactly in terms of incomplete Beta functions, giving
\be\bal
y & = \frac{C}{6 \cN r_0^2} B\left( \frac{z^6}{z^6 +r_0^6 } ; \frac{1}{6}, \frac{1}{3} \right)   \\
A_t & = \frac{\rho \lambda }{24 \pi ^2 \cN\, L^{4}\,  r_0^2} B\left( \frac{z^6}{z^6 +r_0^6 } ; \frac{1}{6}, \frac{1}{3} \right) \ ,
\eal\ee
where 
\be
 r_0^6 = \mathcal{N} ^{-2} \left(\frac{\rho^2\, \lambda}{4\pi^2L^4 }-C^2\right) \ .
\ee
Since $ y = r \sin\theta $, its asymptotic value coincides with $\mtilde$ defined in Eq.~\eqref{eq:D3D7_quark_mass_condensate}, $y(\infty) = \mtilde = \mass\,(2\pi)\, L^2/{\sqrt{\lambda}}$. Additionally, the asymptotic value of $A_t$ is again the chemical potential, $\mu=A_t(\infty)$. Using the asymptotic expansion of the solutions, we can then solve $\rho$ and $C$ in terms of the quark mass and chemical potential
\be
\label{eq:D3D7density_and_C_low_temp}
\bal
\rho & = \frac{16\pi}{\lambda^2} \, {\cN} \, \gamma_* \, \mu \left( \mu^2 -  M_\mt{q}^2\right)   \\
C & =  {\cN} \left(\frac{2\pi}{\sqrt{\lambda}}\right)^3 \gamma_* \, M_\mt{q}\, \left( \mu^2 - M_\mt{q}^2\right)  \ , \\
\eal\ee
where $\gamma_* = (B(1/6,1/3)/6)^{-3}\approx 0.363$. 

The analytic solution can be expanded about the origin, which will be crucial to understand the low-temperature behavior of the entropy
\be\label{eq.y_and_z_smallz}\bal
y & = C \, \frac{z}{r_0^3 \, \cN} + {\cal O}(z^7)  \\
A_t & = \frac{\lambda \rho}{4\pi^2 L^4} \frac{z}{r_0^3 \, \cN} + {\cal O}(z^7) \ .
\eal\ee
In particular, at $z=0$,
\begin{equation} \label{eq.angle_horizon}
\tan{\theta_\mt{h}} = \frac{C}{r_0^3\cN}  = \frac{\mass}{\sqrt{\mu^2-\mass^2}}\quad \Rightarrow \quad  \cos{\theta_\mt{h}} = \frac{1}{\mu} \sqrt{\mu^2 + \mass^2} \ ,
\end{equation}
using which we obtain the contribution of the flavor sector to the entropy,
\begin{equation}
s_{\text{\tiny{f}}}\, (T=0) \, =  \,   \pi L^2 \, \tilde{\LL}\, \Big|_{r\, =\, 0} =\cos\theta_\mt{h}\,  \tilde{\tilde\LL}\, \Big|_{z\, =\, 0} = \frac{\Nf \Nc \gamma_* }{2\sqrt{\lambda}}\ \mu(\mu^2-\mass^2)\,.
\end{equation}
Note that the change of radial coordinate from $r$ to $z$ is responsible for the appearance of $\cos\theta_\mt{h}$ in the latter expression.

\begin{figure}[t]
    \centering
    \includegraphics[width=.47\textwidth]{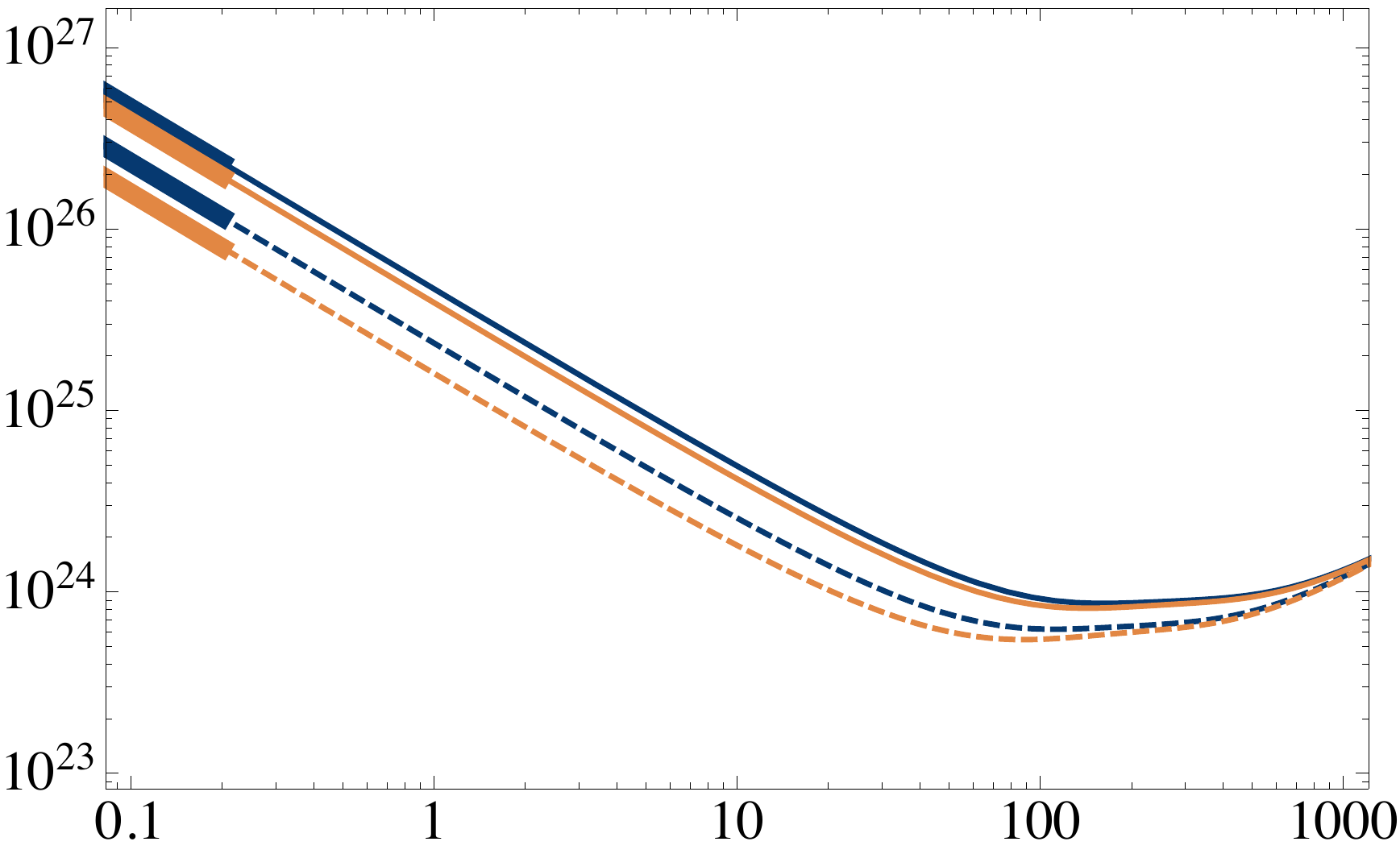}\hfill
    \put(-195,121){$\sigma^{xx}$ [s$^{-1}$]}
	\put(-50,16){$T$ [MeV]}
	\hfill
    \includegraphics[width=.47\textwidth]{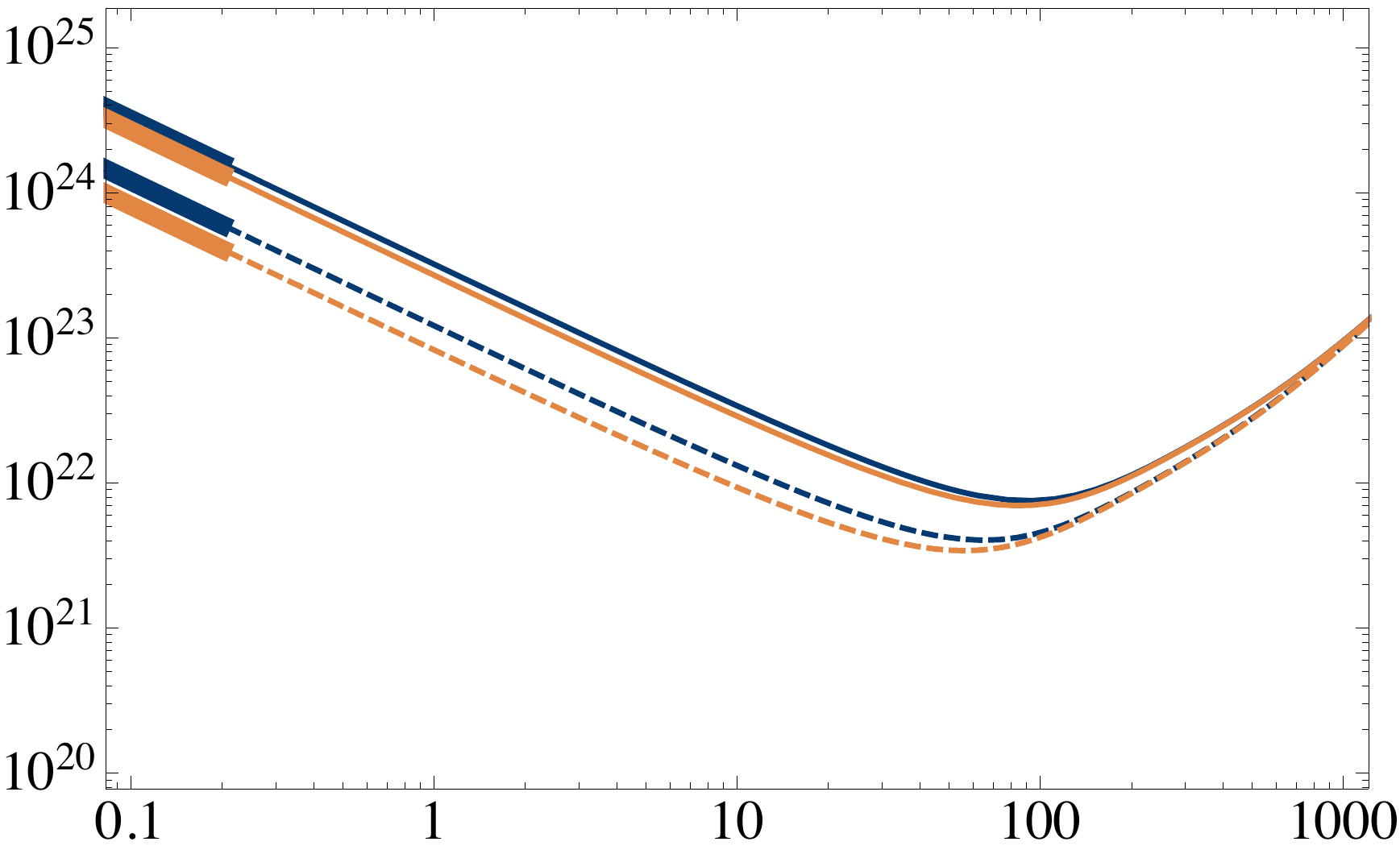}    \put(-195,121){$\kappa^{xx}$ [erg cm$^{-1}\, $K$^{-1}\, $s$^{-1}$]}
	\put(-50,16){$T$ [MeV]}
    \caption{\small Electrical (left) and thermal (right) conductivites as functions of temperature in the D3-D7 system in the probe approximation, for $\mass=210.76$ MeV and $\mu = 450$ MeV ($\mu = 600$ MeV), corresponding to the thick orange (dashed blue) curve. At low temperatures, the constants predicted by Eq.~\eqref{eq:D3D7density_and_C_low_temp}, represented by the thicker solid lines, are recovered.}
    \label{fig:ConductivitiesD3D7}
\end{figure}

\begin{figure}[t]
    \centering
    \includegraphics[width=.47\textwidth]{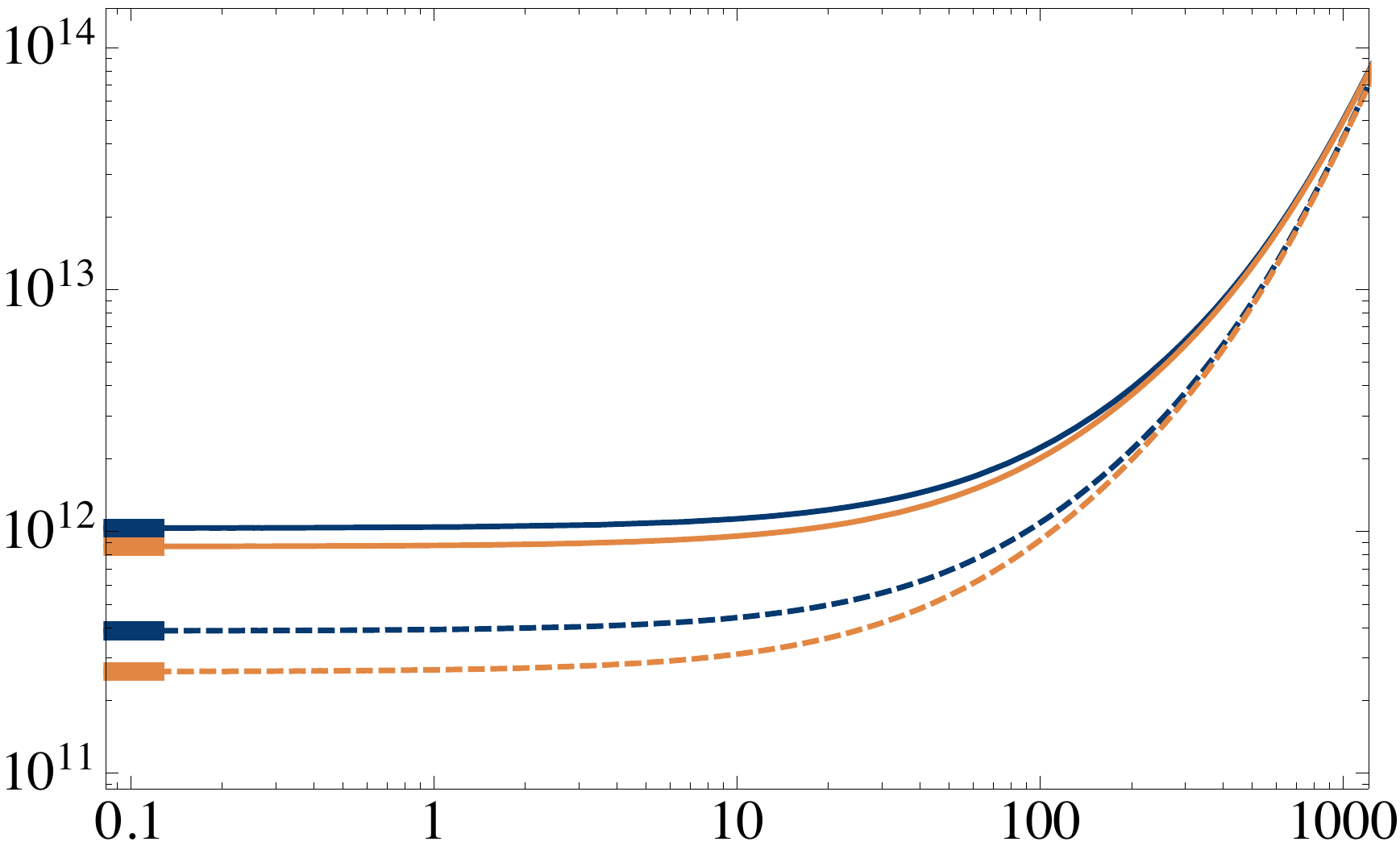}\hfill
    \put(-195,120){$\eta_{\text{\tiny f}}$ [g cm$^{-1}$s$^{-1}$]}
	\put(-50,15){$T$ [MeV]}
	\hfill
    \includegraphics[width=.47\textwidth]{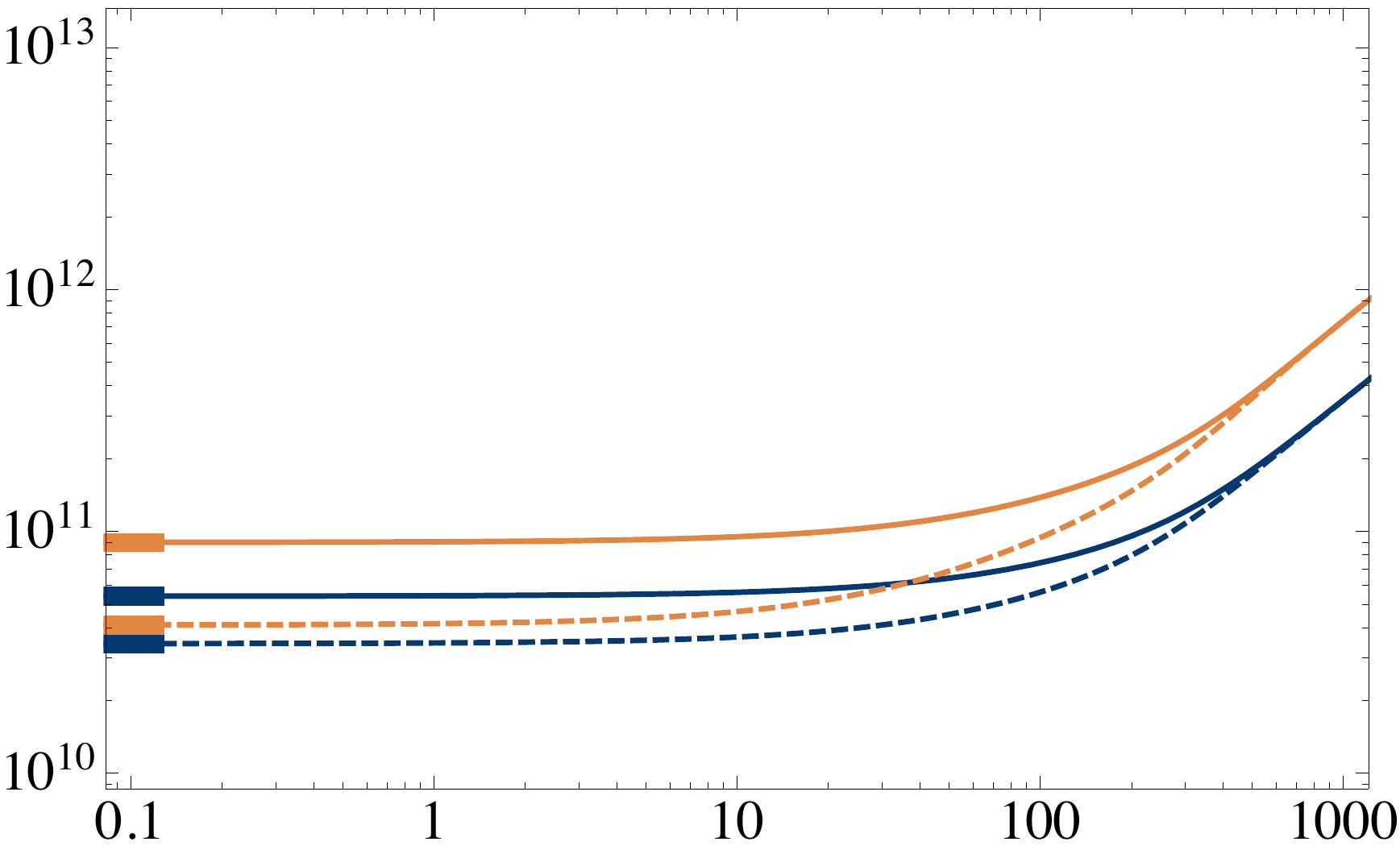}    
    \put(-195,120){$\zeta$ [g cm$^{-1}$s$^{-1}$]}
	\put(-50,15){$T$ [MeV]}
    \caption{\small Shear (left) and bulk (right) viscosities as a function of the temperature in the D3-D7 system in the probe approximation. The blue (orange) curves correspond to the choice $\mass=210.76$ MeV ($\mass=308.55$ MeV).
    We picked two distinct valued for the chemical potential, $\mu = 450$ MeV and $\mu = 600$ MeV; corresponding to dashed and solid curves, respectively. At low temperatures, the constants predicted by \eqref{eq:etalowT} and \eqref{eq:zetalowT} as represented by thicker lines are recovered.}
    \label{fig:ViscositiesD3D7}
\end{figure}

With all this information, the behavior of the transport properties at small temperatures can be determined in a  straightforward way. First of all, the contribution of the flavor sector  to the shear viscosity is given by
\begin{equation}\label{eq:etalowT}
    \eta_{\text{\tiny f}} = \frac{s_{\text{\tiny f}}}{4\pi}  = \frac{\Nf \Nc \gamma_* }{8\pi\sqrt{\lambda}}\ \mu(\mu^2-\mass^2) \ .
\end{equation}
In Fig.~\ref{fig:ViscositiesD3D7} (left), we see how the numerical results nicely approach this value in the low-temperature limit.

Regarding the bulk viscosity, note that we have been able to write the value of the angle at the horizon in terms of the chemical potential in Eq.~\eqref{eq.angle_horizon}. With that and the relation between $\mu$ and $\rho$ given by Eq.~\eqref{eq:D3D7density_and_C_low_temp}, we can compute
\be
\rho \frac{\partial \theta(0)}{\partial \rho}\Big|_{M_\mt{q}\text{ constant, }T=0} = -   \frac{M_\mt{q}\, \sqrt{\mu^2-M_\mt{q}^2}}{3 \mu^2 -  M_\mt{q}^2 } \,,
\ee
which can be substituted into the expression for the bulk viscosity
\eqref{eq.BulkViscosityProbe}. Taking into account that the contribution from $s\, \partial\chi_\mt{h}/\partial s$ vanishes in the zero-temperature limit, we obtain the finite result
\be\label{eq:zetalowT}
\zeta = \Nc \Nf \frac{\gamma_*}{2\pi} \frac{M_\mt{q}^2 \mu}{\sqrt{\lambda}} \frac{(\mu^2 - M_\mt{q}^2)^2}{(3\mu^2 - M_\mt{q}^2)^2} \ .
\ee
The behavior of the bulk viscosity at low temperatures is shown in Fig.~\ref{fig:ViscositiesD3D7} (right).

Curiously, the ratio $\zeta/\eta_\mt{f}$, or equivalently $\zeta/s_\mt{f}$, can be written in a very compact form in terms of the speed of sound. This can be seen by analyzing the quasinormal mode dispersion relations of the gravity dual. The relevant fluctuation mode originating from the vector gauge sector is called the zero sound and can be determined analytically at $T=0$ \cite{Kulaxizi:2008kv}, giving
\be\label{eq:zerosound}
 c_s^2 = \frac{\mu^2-M_\mt{q}^2}{3\mu^2-M_\mt{q}^2} \ .
\ee
We note that this equals the stiffness $(\partial p/\partial\varepsilon)|_s$, commonly referred to as the speed of first sound (see Eq. \eqref{eq_app:speedofsoundD3D7} in Appendix~\ref{app:D3D7 computations}) for the D3-D7 probe brane model in the limit $T\to 0$ \cite{Kulaxizi:2008jx,Kim:2008bv}; the same holds more generally for any probe brane intersections \cite{Itsios:2016ffv}. Solving for $(M_\mt{q}/\mu)^2$ from (\ref{eq:zerosound}) yields for the bulk to shear viscosity ratio:
\be\label{eq:zetaetaratio}
 \frac{\zeta}{\eta_\mt{f}} = 6c_s^2\left(\frac{1}{3}-c_s^2\right) \ .
\ee
Note that we are including only the flavor contribution to the shear viscosity, since the glue contribution vanishes at zero temperature. At finite temperature the ratio is highly suppressed $\zeta/\eta_\mt{f} \sim \Nf/\Nc$.

We can compare the above result with the holographic bound for the bulk to shear viscosity ratio $\frac{\zeta}{\eta}\geq 2\left(\frac{1}{3}-c_s^2\right)$ proposed in \cite{Buchel:2007mf} and generalized to finite density in \cite{Gouteraux:2011qh} (fixing $c_s^2$ to be the stiffness). The bound is violated for any speed $c_s^2<1/3$, which according to \eqref{eq:zerosound} is always the case.
This thus provides yet another example \cite{Hoyos:2016cob,Ecker:2017fyh}, where holographic conjectures are violated at finite density \footnote{Violations of the bound can also occur if the field theory is on a curved manifold \cite{Buchel:2011uj}.}.

As already discussed in detail in  \cite{Czajka:2018bod}, the dependence of the viscosity ratio on $1/3-c_s^2$ --- a quantity measuring the deviation from conformal invariance --- turns out to involve a different power in the strong coupling and weak coupling (kinetic theory) calculations (see \cite{Arnold:2006fz} for the latter), although $\zeta/s\propto (1/3-c_s^2)$ turns out to have the same linear dependence in both regimes. It should be noted that this comparison has been made in the low density, high temperature regime, while we are working in the opposite limit of high densities and low temperatures. This does, however, not seem to affect the qualitative dependence of the viscosity ratio on $1/3-c_s^2$.

Finally, let us move on to the analytical expressions of the conductivities at small temperatures. From \eqref{eq.DCconductivity} and the expression we have for the charge density, Eq.~\eqref{eq:D3D7density_and_C_low_temp}, we get
\be\label{eq:cond_low_temp}
 \sigma = \frac{2 \Nc\Nf\gamma_*}{\pi\, \lambda^{\frac{3}{2}}}\, \frac{\mu (\mu^2-\mass^2)}{T^2}\ .
\ee
Recall that the actual conductivity $\sigma^{xx}$ has an additional prefactor, which at leading order at low temperature reads
\begin{equation}
    \frac{s\, T}{\varepsilon+p} \ =\  \frac{\sqrt{\lambda}}{2}\, \frac{T}{\mu}\, +\, \OO(T^2) \ .
\end{equation}
Consequently,  the final result for the conductivity at low temperatures obtains the form
\begin{equation}\label{eq:sigmaxxZT}
    \sigma^{xx}= \frac{ \Nc\, \Nf\, \gamma_*}{\pi\, \lambda}\, \frac{(\mu^2-\mass^2)}{T}\ .
\end{equation}
Again, from \eqref{eq.kappa_equation}, a similar expression can be found for the thermal conductivity
\begin{equation}
    \kappa^{xx} = \frac{ \Nc\, \Nf\, \gamma_*}{2\, \pi\, \sqrt{\lambda}}\, \frac{\mu\, (\mu^2-\mass^2)}{T}\ .
\end{equation}
These low temperature results for the conductivities are included in Fig.~\ref{fig:ConductivitiesD3D7} as the thick asymptotic lines.

Interestingly, at low temperatures we explicitly see that the Wiedemann-Franz law is violated, {\emph{i.e.}}, that the ratio $\frac{\kappa^{xx}}{\sigma^{xx}T}$ is not constant (which would be the Lorenz number):
\be\label{eq:WFlaw}
 \frac{\kappa^{xx}}{\sigma^{xx}T} = \frac{\sqrt\lambda}{2}\frac{\mu}{T} \ .
\ee

\subsection{V-QCD}\label{app:resultsVQCD1}

Next, we move on to discussing the results from the V-QCD setup. This time, we begin from the limit of small temperatures, which can be handled analytically, moving only thereafter to the more generic numerical results.

\subsubsection{V-QCD: small-$T$ limit as transseries}\label{app:resultsVQCD}

It is also possible to derive some analytic results for the transport coefficients in the quark matter phase of V-QCD in the limit of zero temperature. The methods and results differ, however, quite a bit from the D3-D7 computation of the previous subsection. We discuss them in detail below and in Appendix~\ref{app:ads2}. 

It is useful to first recall the phase diagram of V-QCD at low temperatures (and zero quark masses). At small $\mu$, the dominant phase is the confined chirally broken phase, dual to the horizonless thermal gas background with nonzero tachyon condensate, whereas at high $\mu$ the dominant phase is the deconfined, chirally symmetric quark matter phase. Since we are not considering implementations of nuclear matter in this article, we take the model exactly as defined in~\eqref{eq:ActionDefsFlavor}. Then the phase transition between the thermal gas and quark matter phases takes place at $\mu \approx 394$, $406$, and $428$~MeV (at $T=0$) for the potentials {\bf{5b}}, {\bf{7a}}, and {\bf{8b}}, given in Appendix~\ref{app:vqcd}. The non-Abelian nuclear matter configurations of~\cite{Ishii:2019gta} appear at intermediate chemical potentials. If the configurations are matched with traditional models of nuclear matter at low density as explained in~\cite{Ecker:2019xrw}, the nuclear matter phase ranges from $\mu \approx 310$~MeV to around 500~MeV, with the exact number depending on the details of the matching. 

One should recall that at very low temperatures in the quark matter phase, one expects to witness the pairing of quarks, which is absent in our model. The low-temperature asymptotics may in any case give a reasonable description of transport above the possible pairing transition. It also turns out to have an interesting structure determined by an AdS$_2$ fixed point, which leads to the low-temperature asymptotics of various observables being described in terms of transseries.

The IR geometry in the quark matter phase is $AdS_2 \times \mathbb{R}^3$ at zero temperature, signaling the presence of a quantum critical line~\cite{Alho:2013hsa}. We discuss the asymptotic geometry of this phase in detail in Appendix~\ref{app:ads2} and present here only the main results. The flow to the IR fixed point, which gives the asymptotically AdS$_2$ geometry at zero temperature, is found in terms of a transseries. That is, the flow of the dilaton and the metric can be written in the form
\be
 F(r) = \sum_{i,j=0}^{\infty} F_{ij} \hat r^{i} \hat r^{\alpha j} \ ,
\ee 
where $F$ represents a generic background function, $r_0-r = \hat r$ is the distance from the end of space at $r=r_0$, and $\alpha$ is a positive parameter that can be computed from the action (see Appendix~\ref{app:ads2} for details). For the three potentials used in this article  we find that
\begin{align} \label{eq:alphavals}
    \alpha &\approx 0.3558 \ ,& \quad &(\textrm{potentials 5b})\ ; &\nonumber \\
    \alpha &\approx 0.6566\ ,& \quad &(\textrm{potentials 7a})\ ; &\\
    \alpha &\approx 0.4850\ ,& \quad &(\textrm{potentials 8b})\ . &\nonumber 
\end{align}

By using the transseries, the thermodynamics at low temperatures in the (unpaired) quark matter phase can be analyzed. This is done by modifying the geometry near the AdS$_2$ end through adding a blackening factor (see Appendix~\ref{sec:thermoVQCD}). First we find that the entropy goes to a constant at zero temperature, with the leading corrections decaying as $T^{2\alpha}$ or $T$ if $\alpha$ is below or above $1/2$, respectively. For the viscosities and conductivities we find that
\begin{align} \label{eq:vqcdviscsmallT}
    \eta &\sim T^0 \ , & \quad \zeta &\sim \left\{\begin{array}{cc}
        T & \quad \mathrm{if} \ \ 0 <\alpha<1/2  \\
        T^{2\alpha} & \quad \mathrm{if}\ \  1/2 <\alpha<1
    \end{array} \ , \right. & \\
    \sigma^{xx} &\sim T\ , & \quad \kappa^{xx} &\sim T& 
    \label{eq:vqcdcondssmallT}
\end{align}
as $T \to 0$, with the subleading corrections to $\eta$ behaving in the same way as the corrections to entropy.  Note that the Wiedemann-Franz law is not obeyed, as $\kappa^{xx}/(\sigma^{xx}T)\sim T^{-1}$, similarly to the D3-D7 case of Eq.~(\ref{eq:WFlaw}).

\begin{figure}[!t]
\begin{center}
\includegraphics[width=0.8\textwidth]{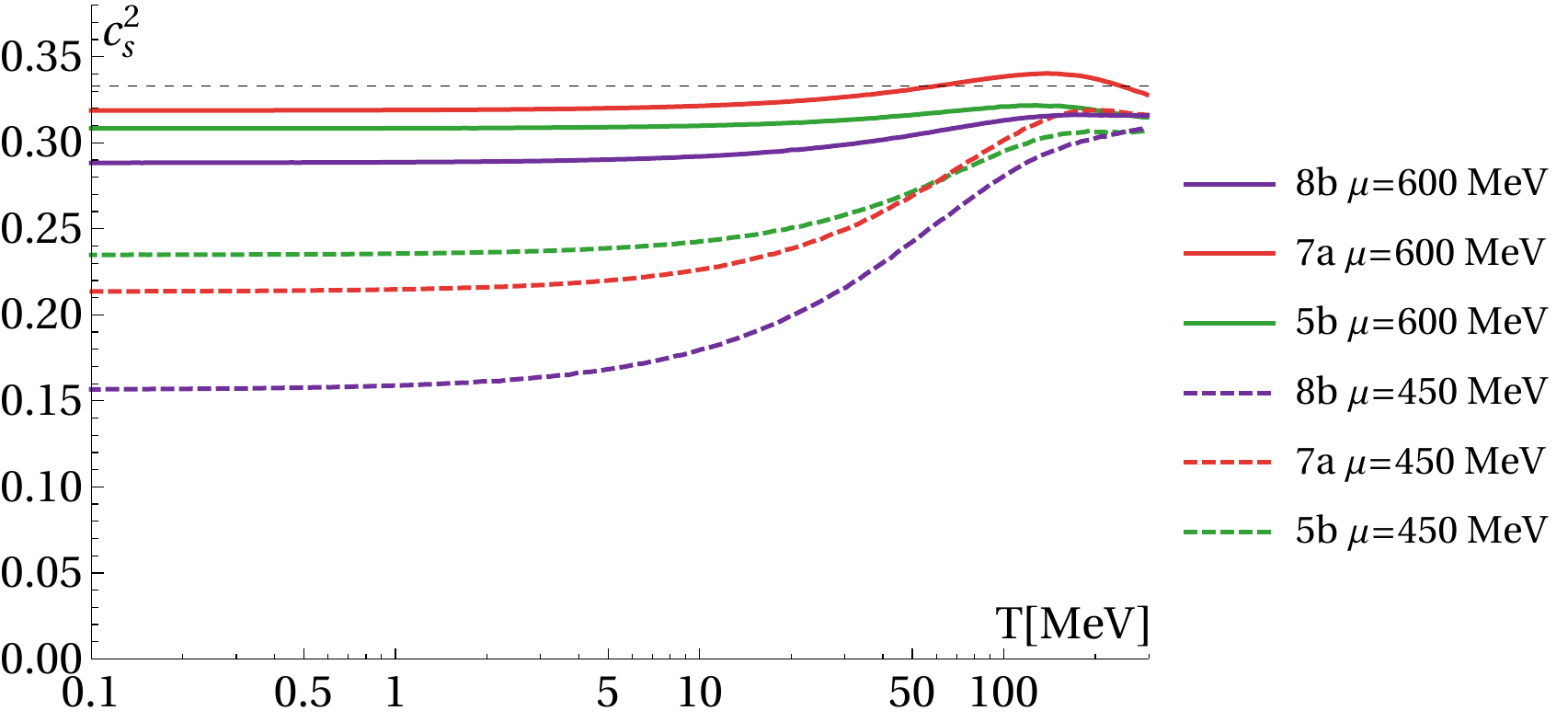}
\end{center}
\caption{The speed of first sound squared in V-QCD. The green, red, and violet curves correspond to the potentials \textbf{5b}, \textbf{7a}, and \textbf{8b}, respectively, whlie the dashed and solid curves stand for $\mu=450$~MeV and at $\mu=600$~MeV, respectively. The dashed thin black line shows the value $c_s^2=1/3$ obtained in conformal theories.}
\label{fig.VQCDcs2}
\end{figure}

\begin{figure}[!t]
\begin{center}
    \includegraphics[width=.47\textwidth]{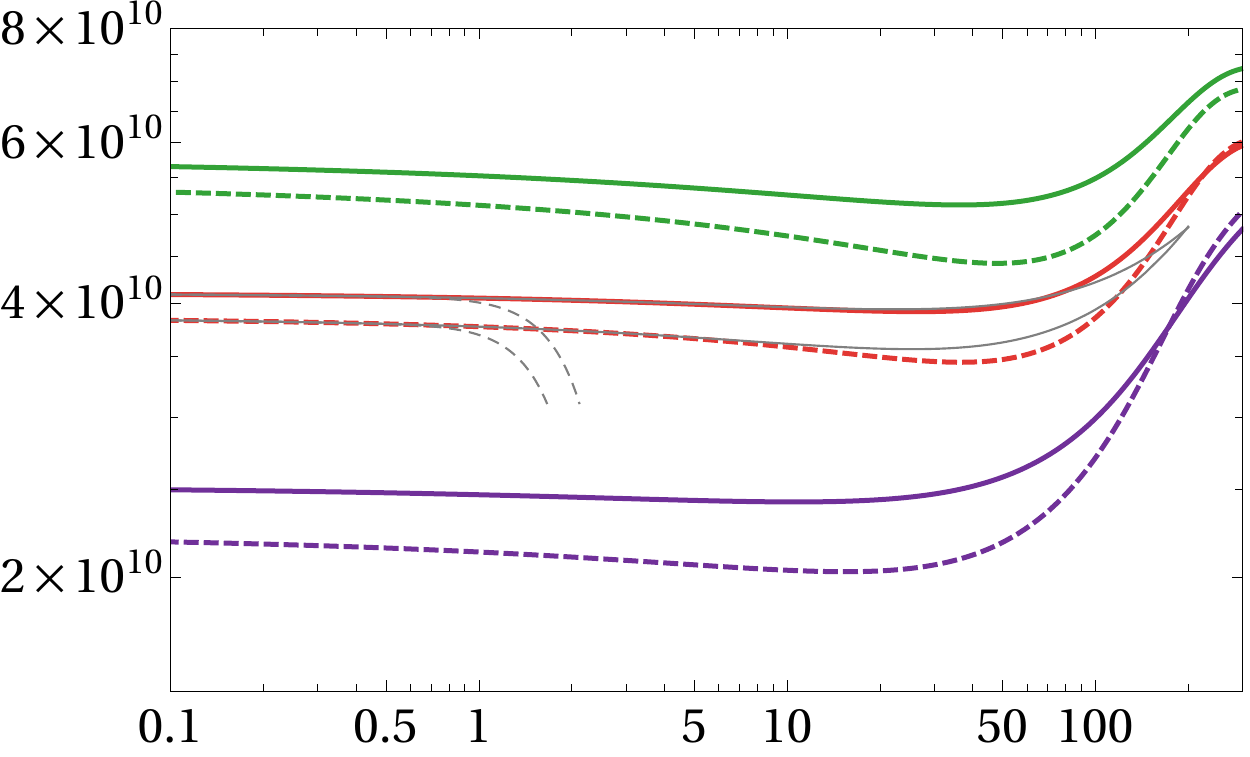}\hfill
    \put(-185,117){$\sigma^{xx}/T$ [s$^{-1}$K$^{-1}$]}
	\put(-45,17){$T$ [MeV]}
	\hfill
    \includegraphics[width=.47\textwidth]{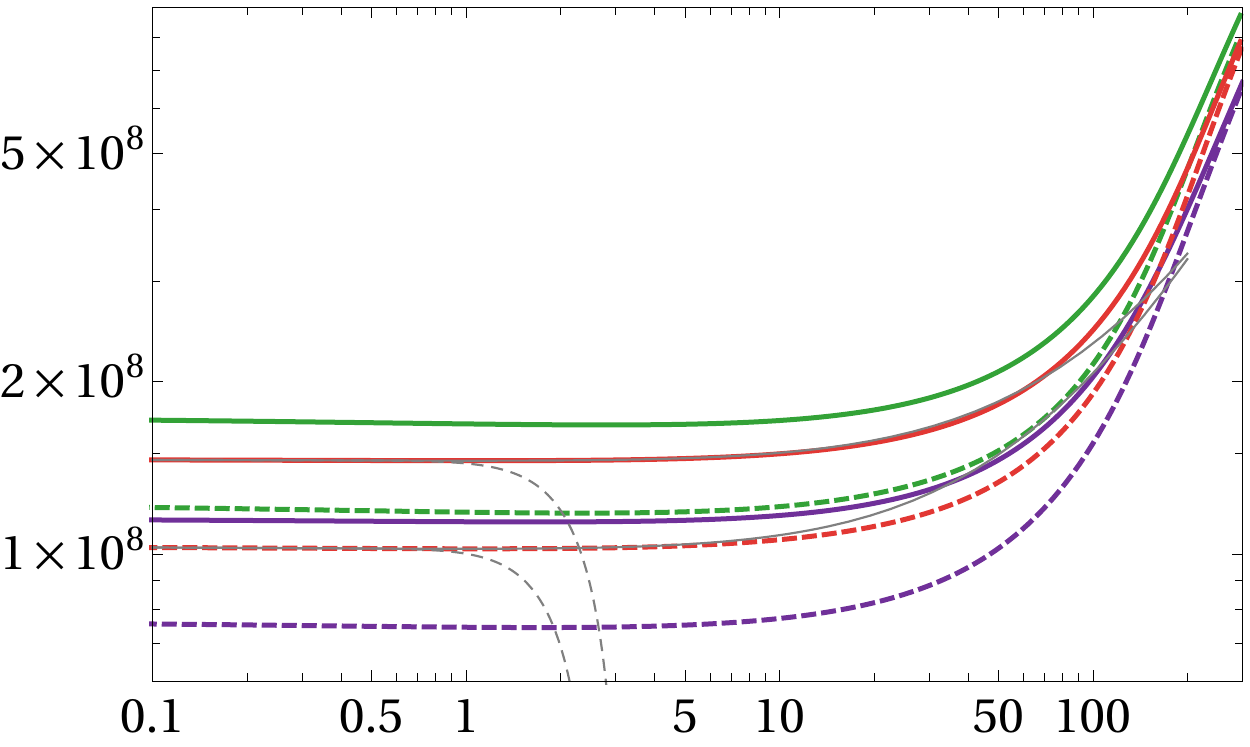}    
    \put(-185,117){$\kappa^{xx}/T$ [erg cm$^{-1}$s$^{-1}$K$^{-2}$]}
	\put(-45,17){$T$ [MeV]}
\end{center}
\caption{The electrical and thermal conductivities in V-QCD.  The various colored curves show the data for $\mu=450$ or $600$~MeV and different potentials with the same notation as in Fig.~\protect\ref{fig.VQCDcs2}. The thin dashed and solid gray curves are polynomial and transseries fits, respectively, to the data for potentials \textbf{7a}.
} 
\label{fig.VQCDconds}
\end{figure}

\subsubsection{V-QCD: numerical results}\label{sec:resultsVQCD}

Next, we proceed to study the transport coefficients in V-QCD at higher temperatures. In this case, they are extracted from the numerical solution of the background using the formulae of Secs.~\ref{sec:conductivities} and~\ref{sec:viscosities}. Specifically, for the conductivities we use Eq.~\eqref{eq.DCconductivity}, while the shear viscosity is simply given by $\eta = s/(4\pi)$ with the entropy computed form the area of the black hole. The bulk viscosity is finally found by using the Eling-Oz formula~\eqref{eq.BulkViscosityFinal} or from the Kubo formula with input from the fluctuations of the metric in the helicity zero sector (see, {\emph{e.g.}}, Ref.~\cite{Gubser:2008sz} and Appendix~\ref{app:finiteom}). In our work, we carried out both computations and checked that the results agree.

Before dwelling on the transport properties, we briefly display results for the adiabatic speed of first sound $c_s^2=\left(\partial p/\partial\varepsilon\right)_s$ as a function of temperature for the three choices of potentials at $\mu=450$~MeV and at $\mu=600$~MeV. As can be seen from  Fig.~\ref{fig.VQCDcs2}, the speed of sound increases with $T$ at small temperatures, and typically reaches a maximum between $T=100$ and $T=200$~MeV. For the potential \textbf{7a} at $\mu=600$~MeV, the speed of sound exceeds the conformal value $c_s=1/\sqrt{3}$. Upon closer inspection, it can be seen that the speed of sound keeps increasing with the chemical potential in all cases, and the conformal value is exceeded for all three potentials at higher values of $\mu$ (not shown in the figure) even for $T=0$. 

The electrical and thermal conductivities for V-QCD are in turn displayed in Fig.~\ref{fig.VQCDconds}. We have normalized the results by using the expected linear behavior at small temperatures from~\eqref{eq:vqcdcondssmallT}, which is well reproduced by the data.

\begin{figure}[!t]
\begin{center}
    \includegraphics[width=.47\textwidth]{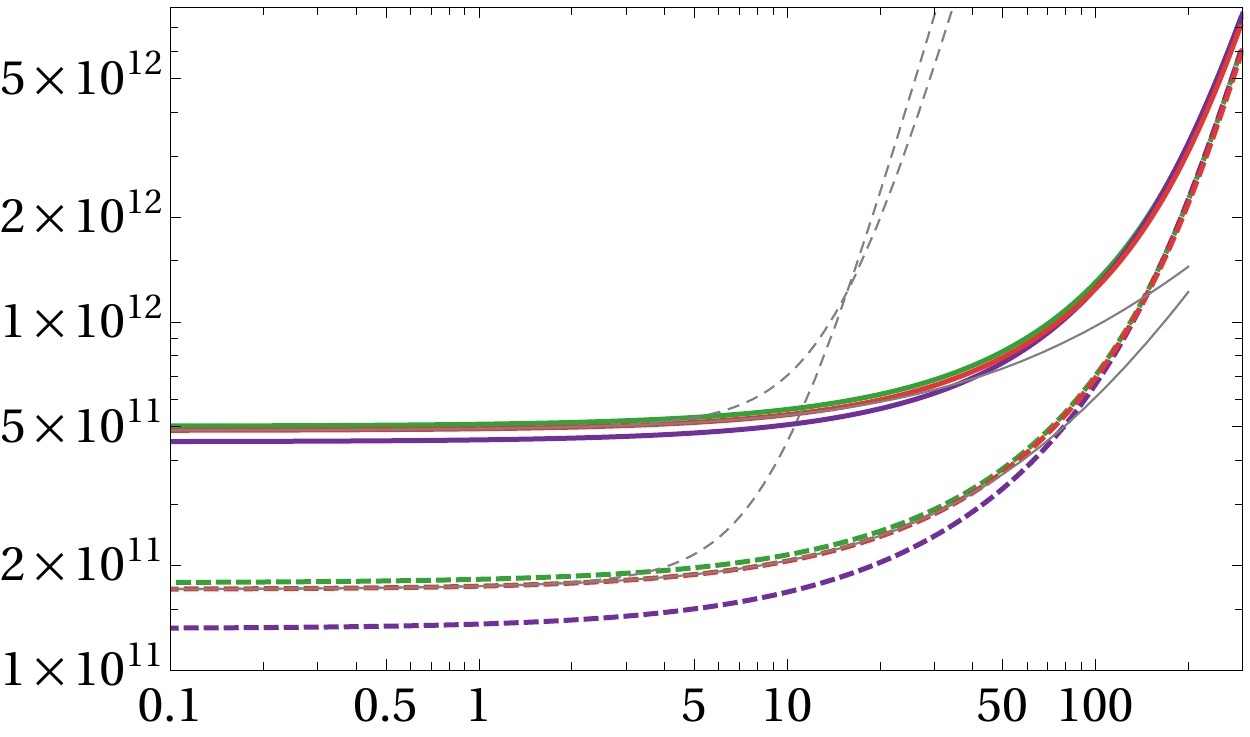}\hfill
    \put(-185,117){$\eta$ [g cm$^{-1}$s$^{-1}$]}
	\put(-45,17){$T$ [MeV]}
	\hfill
    \includegraphics[width=.48\textwidth]{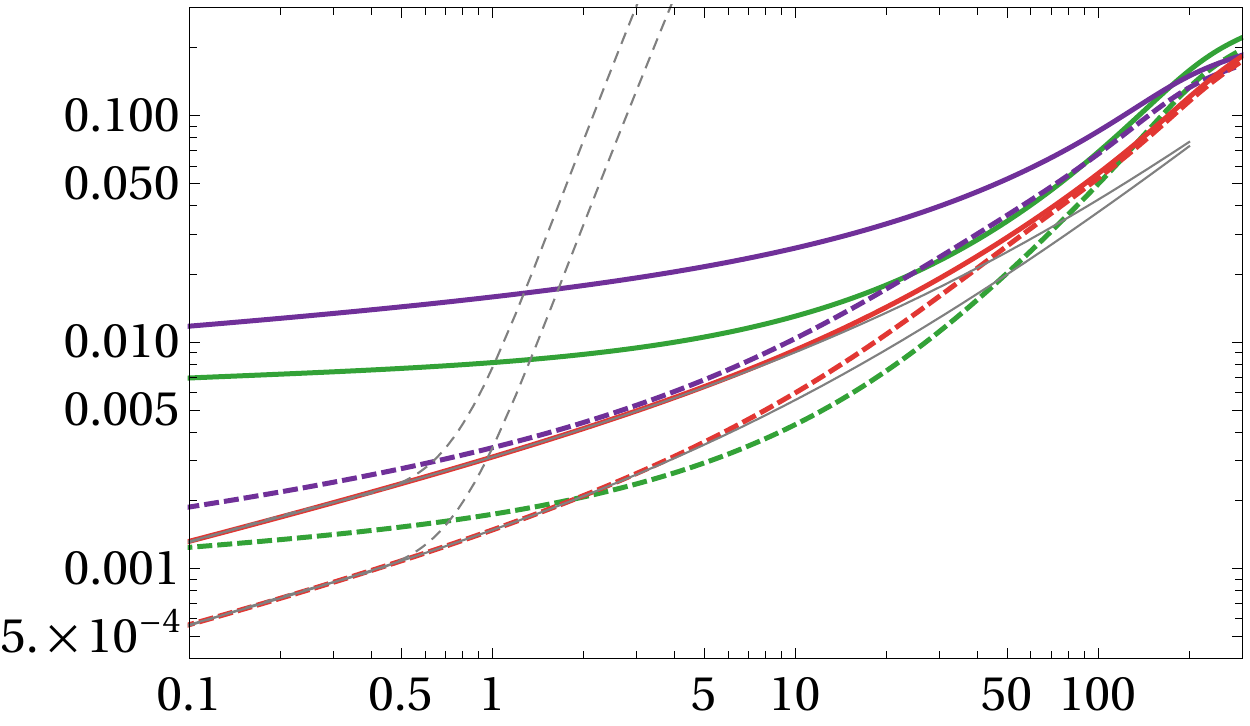}    
    \put(-185,117){$\zeta/T$ [g cm$^{-1}$s$^{-1}$K$^{-1}$]}
	\put(-45,17){$T$ [MeV]}
\end{center}
\caption{The shear and bulk viscosities in V-QCD. 
The notation follows that of Figs.~\protect\ref{fig.VQCDcs2} and~\protect\ref{fig.VQCDconds}.
}
\label{fig.VQCDviscs}
\end{figure}

In Fig.~\ref{fig.VQCDviscs}, we then show the shear and bulk viscosities as functions of $T$ for the same values of chemical potentials we inspected above. Note that we have here divided the bulk viscosity by the temperature in order to more clearly display the details of this quantity. The thin solid and dashed gray curves are respectively fits obtained using the transserries (including powers of the from $T^{i+\alpha j}$ with integer $i$ and $j$) and a naive Taylor expansion around $T=0$. For all curves, we have fitted four free parameters  and used data for the potential \textbf{7a} with $T \lesssim 0.5$~MeV. As one can readily observe, the transseries fit describes the data considerably better than the Taylor fit, which strongly suggests that its asymptotics are indeed given by a transseries as we argue in Appendix~\ref{app:ads2}. The leading low-temperature asymptotics of $\zeta$ furthermore agrees with the expectation from Eq. ~\eqref{eq:vqcdviscsmallT} for the potentials \textbf{5b} and \textbf{7a}. For the potential \textbf{8b} the asymptotic behavior is less clear --- this is apparently the case because the value of $\alpha$ in~\eqref{eq:alphavals} is quite close to the critical value $\alpha=1/2$. Similarly to the D3-D7 model, the Buchel bound for the bulk to shear viscosity ratio would be violated if one uses $c_s^2$ (depicted in Fig.~\ref{fig.VQCDcs2}) to test it. 

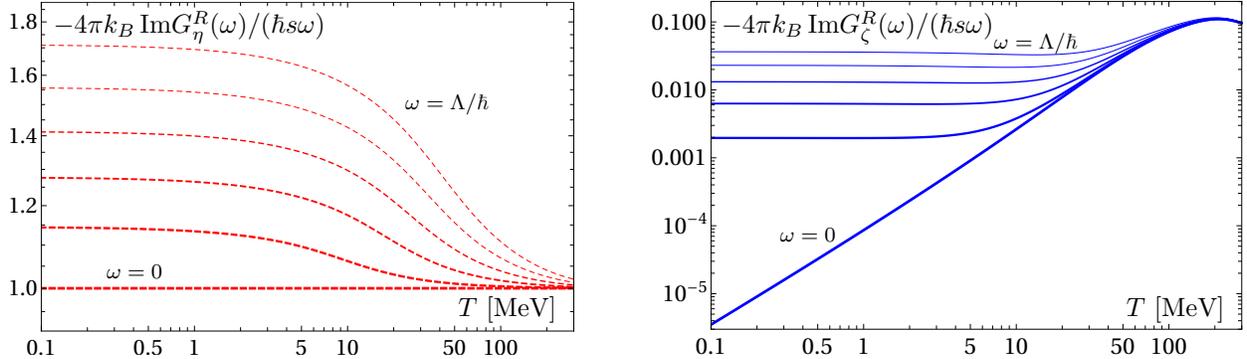
\begin{figure}[!t]
\begin{center}
    \includegraphics[width=.46\textwidth]{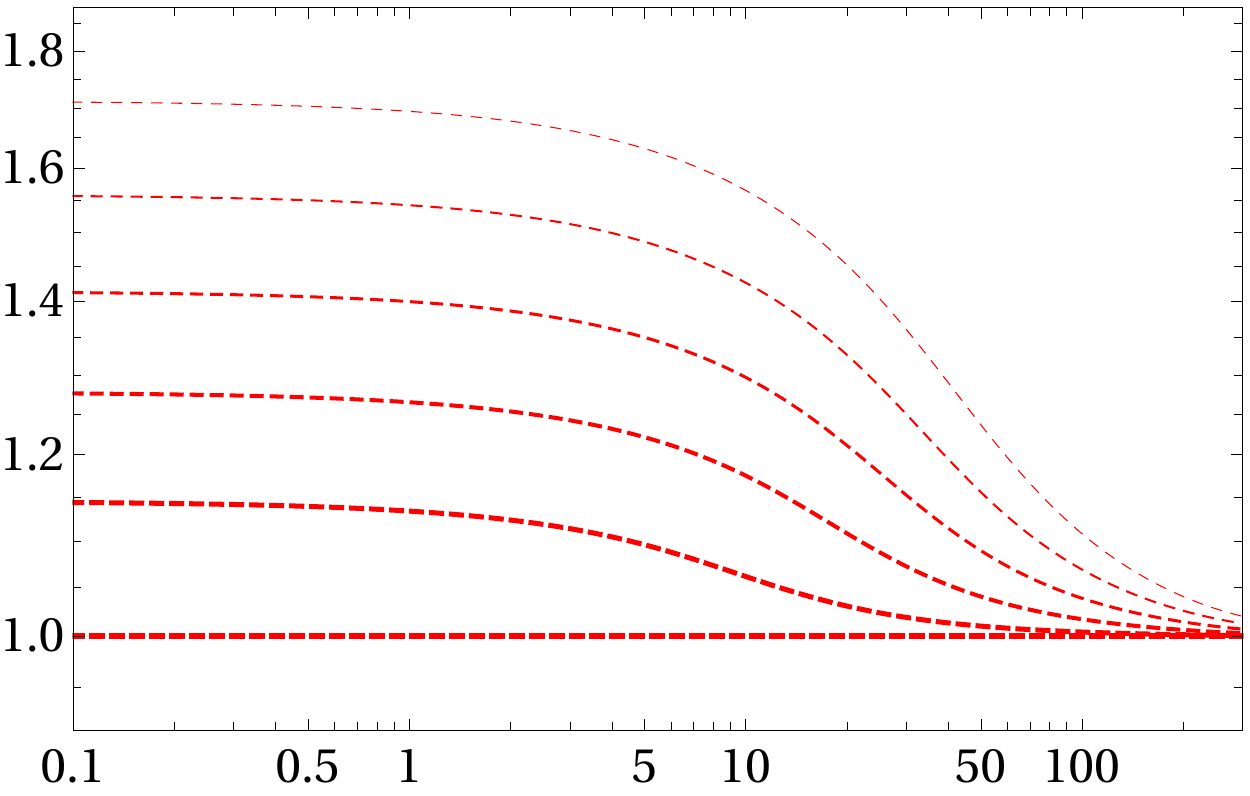}\hfill
    \put(-198,124){$-4\pi k_B\, \mathrm{Im}G_\eta^R(\omega)/(\hbar s\omega)$}
	\put(-45,17){$T$ [MeV]}
	\put(-178,31){\scriptsize$\omega = 0$}
	\put(-65,95){\scriptsize$\omega = \Lambda/\hbar$}
	\hfill
    \includegraphics[width=.48\textwidth]{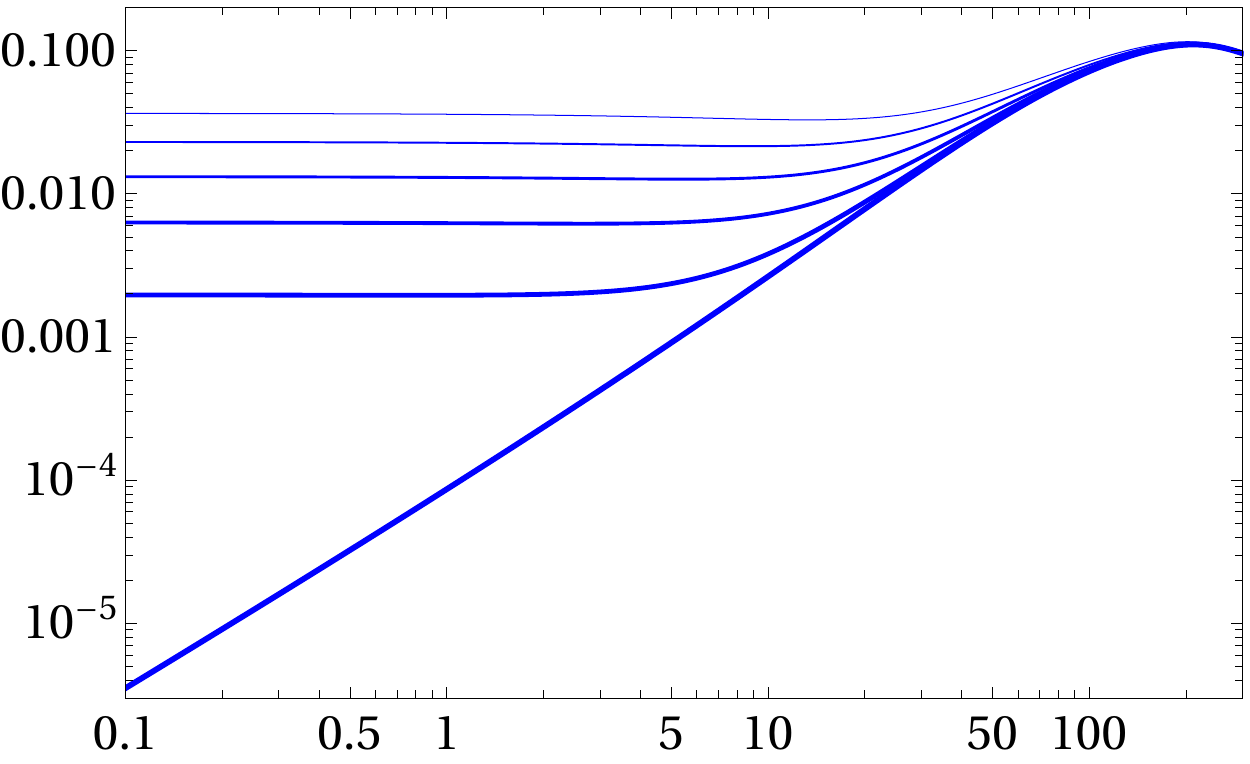}    
    \put(-198,124){$-4\pi k_B\,\mathrm{Im}G_\zeta^R(\omega)/(\hbar s\omega)$}
	\put(-45,17){$T$ [MeV]}
	\put(-176,45){\scriptsize$\omega = 0$}
	\put(-95,118){\scriptsize$\omega = \Lambda/\hbar$}
\end{center}
\caption{The dependence of the viscosities on $\omega$ in V-QCD with potentials \textbf{7a}. We show the ratios of the correlators related to shear (left plot) and bulk (right plot) 
viscosities to the ``standard'' value $\hbar s/(4\pi k_B)$ at $\mu=500$~MeV as a function of $T$. The values of $\omega$ range from $0$ (thick curves) to $\Lambda/\hbar$ (thin curves) in steps of $0.2 \Lambda/\hbar$, where $\Lambda = 211$~MeV is the characteristic scale of the model.}
\label{fig.VQCDviscratioomega}
\end{figure}

In Fig.~\ref{fig.VQCDviscratioomega} we finally plot the imaginary values of the retarded correlators $G_\eta^R(\omega)$ and  $G_\zeta^R(\omega)$, which generalize the shear and bulk viscosities to finite frequency as suggested by the Kubo formulas
\be
 \eta = - \lim_{\omega \to 0} \frac{\mathrm{Im}G_\eta^R(\omega)}{\omega} \ , \qquad  \zeta = - \lim_{\omega \to 0} \frac{\mathrm{Im}G_\zeta^R(\omega)}{\omega} \ ;
\ee
see Appendix~\ref{app:finiteom} for details. We normalized the results using the entropy density (or $\hbar s/(4\pi k_B)$ to be precise) such that the normalized shear viscosity equals unity at $\omega=0$. The dependence of the results on the frequency is strong at small temperatures. Notice that the bulk viscosity no longer vanishes as $T \to 0$, if the frequency is nonzero, signalling a breaking of conformal invariance.

\section{Discussion}\label{sec:discussion}

Holography has proven to be a very useful tool in the description of dense fundamental matter, purportedly strongly interacting in the environment provided by neutron star cores. The field has matured to a point where the predicted equilibrium properties of dense QCD matter are consistent with known physics and observations across the phase diagram, from small densities to high, hot and cold, so it is time to move on to quantities more challenging to traditional field theory methods. In this paper, we have done this by extracting holographic lessons for dynamical dense systems in the regime of linear response. In particular, we have performed a detailed study of the transport properties of unpaired quark matter in a state of flux realizable inside massive neutron stars \cite{Annala:2019puf}.

In our work, we found strongly coupled holographic quark matter to respond to perturbations in a way that dramatically differs from what one might have expected based on naive extrapolations of weak-coupling results; for a detailed discussion of this point, see also our previous article \cite{Hoyos:2020hmq}. Note that we have only considered what is commonly referred to as the unpaired quark matter phase, where quark pairing does not occur in any form. We believe, however, that the existence of the said discrepancies is independent of this detail, and that similar differences will be found in various ``color superconducting" phases once such results become available. The key distinction between the weak- and strong-coupling approaches namely stems from the fact that at finite density and strong coupling, there is no quasiparticle description of the system, and hence the mere existence of Fermi surfaces that play a pivotal role in perturbative studies is unclear. Conceptually, and also technically, a more straightforward extension of our investigation would be to allow for non-zero magnetic fields in the unpaired phase. This would open up an avenue for addressing the time evolution of magnetic fields inside NSs and in particular the coupling to their thermal evolution as required by magnetohydrodynamical simulations of NS mergers. In this case, many new transport coefficients would appear due to parity breaking, also  accessible by our methods.

At lower densities, below those needed to excite strange quarks in the medium, various forms of bound states need to be taken into account. It is currently unknown, whether the nuclear and quark matter phases are separated by a discontinuous first-order transition, and if yes, whether mixed phases are possible. Nevertheless, it is likely that nucleons, too, are strongly interacting at densities close to the phase transition, and a description based on treating their interactions perturbatively becomes invalid. After several years of hiatus since the early works of \cite{Bergman:2007wp,Rozali:2007rx,Kim:2007zm}, a holographic approach to model dense baryonic matter has recently regained some momentum \cite{Ishii:2019gta,Kovensky:2021ddl}, leading, {\emph{e.g.}}, to the development of a unified framework describing both the nuclear and quark matter phases  \cite{Jokela:2020piw} as well as a mixed phase \cite{deBoer:2012ij,BitaghsirFadafan:2018uzs,Kovensky:2020xif} resembling quarkyonic matter \cite{McLerran:2018hbz}. To this end, we find it important that a holographic investigation of transport phenomena be launched also in the confined phase of QCD.

While we have here considered a fairly general holographic setup, there are several models of quark matter that one might equally well employ. These include a phenomenologically deformed D3-D7 model \cite{Fadafa:2019euu,BitaghsirFadafan:2020otb}; the well-known Sakai-Sugimoto model \cite{Rebhan:2014rxa}, consisting of pairs of D8-branes in a Witten background; any non-DBI type actions such as Einstein-Maxwell-scalar configurations; and backreacted brane configurations in the smearing approximation \cite{Nunez:2010sf}. Out of the latter three, the smeared configurations have the undesirable feature that the matter sector is sensitive to the lower-dimensional induced metric, and it is not always easy to find a dimensionally truncated effective action. Similarly, in the Sakai-Sugimoto model the low-temperature phase is confining, so that without considering the backreaction of the flavor D8-branes there is no controlled approximation that would allow the description of deconfined quark matter at low temperatures. Instead, one typically extrapolates the high-temperature deconfined phase down to small temperatures, ignoring the deconfinement transition that is dual to the Hawking-Page phase transition of the background geometry. In addition to this, another complication in the Sakai-Sugimoto context is linked to the fact that the resulting phase is not translationally invariant but subject to the decay to an inhomogeneous ground state triggered by non-trivial Wess-Zumino terms in the brane action \cite{Nakamura:2009tf,Bergman:2011rf}. Numerical techniques for solving the resulting partial differential equations involving square root actions \cite{Jokela:2014dba,Rai:2019qxf} and even for the extraction of transport coefficients therein \cite{Jokela:2016xuy,Jokela:2017ltu} have been developed, but their implementation in the Sakai-Sugimoto context will likely be somewhat tedious. Finally, Einstein-Maxwell-scalar theories have also not been extensively considered at temperatures relevant for NS matter (for a notable exception, see \cite{Mamani:2020pks}), and we hope that our exploration encourages further studies along this path.


\begin{acknowledgments}
We thank Umut G\"ursoy and Elias Kiritsis for useful discussions. C.~H. has been partially supported by the Spanish {\emph{Ministerio de  Ciencia, Innovaci\'on y Universidades}} through the grant PGC2018-096894-B-100. N.~J. and A.~V. have been supported by the Academy of Finland grants no.~1322307 and 1322507 as well as by the European Research Council, grant no.~725369. The work of M.~J. has been supported in part by an appointment to the JRG Program at the APCTP through the Science and Technology Promotion Fund and Lottery Fund of the Korean Government. M.~J. has also been supported by the Korean Local Governments -- Gyeongsangbuk-do Province and Pohang City -- and by the National Research Foundation of Korea (NRF) funded by the Korean government (MSIT) (grant number 2021R1A2C1010834). J.~G.~S. has been supported by grants SGR-2017-754 and PID2019-105614GB-C22, and also acknowledges financial support from the State Agency for Research of the Spanish Ministry of Science and Innovation through the “Unit of Excellence Mar\'ia de Maeztu 2020-2023” award to the Institute of Cosmos Sciences (CEX2019-000918-M) and from the FPU program (FPU15/02551 and EST18/00331).  J.~T. would like to thank the University of Oviedo for providing support during the initial stages of this work under the Grant for Visiting Docent Personnel funded by the Banco Santander. We also acknowledge support from CNRS through the PICS program as well as from the Jenny and Antti Wihuri Foundation.
\end{acknowledgments}


\appendix

\section{Equations of motion}\label{app.eoms}

In this first Appendix, we reproduce the equations of motion for our generic holographic setup discussed in Sec.~\ref{sec:models}. Using Eq.~\eqref{eq.GammaMatrix}, the equations of motion that are derived from \eqref{eq.ActionWithTachyon} read
\be\label{eq.EquationsOfMotion}
\begin{aligned}
& R^{\mu\nu} - \frac{1}{2} g^{\mu\nu}R - \frac{1}{2} \partial^\mu\phi \, \partial^\nu\phi + \frac{1}{4} g^{\mu\nu} \partial_\rho \phi \, \partial^\rho \phi + \frac{1}{2} g^{\mu\nu} \, \pot + \frac{1}{2} \tens\, \Z \, \frac{\sqrt{-\Gamma}}{\sqrt{-g}} \, \Gamma^{(\mu\nu)}  = 0  \\
& \frac{1}{\sqrt{-g}} \partial_\mu \left( \sqrt{-g}\, g^{\mu\nu} \partial_\nu \phi \right) - \partial_\phi \pot  - \frac{\tens \,\Z}{2} \frac{\sqrt{-\Gamma}}{\sqrt{-g}} \left(2 \frac{\partial_\phi \Z}{\Z} + \Gamma^{\mu\nu} \left( \partial_\phi \k \, \partial_\mu \chi \partial_\nu\chi - \partial_\phi \W \, F_{\mu\nu} \right) \right)  = 0  \\
& \frac{1}{ \Z \sqrt{-\Gamma}} \partial_\mu \left( \sqrt{-\Gamma} \, \k\,\Z\, \Gamma^{(\mu\nu)} \partial_\nu \chi \right) - \frac{\sqrt{-g}}{\sqrt{-\Gamma}} \frac{\partial_\chi \pot}{\tens\,  \Z}  - \frac{1}{2}\left(2 \frac{\partial_\chi \Z}{\Z} + \Gamma^{\mu\nu} \left( \partial_\chi \k \, \partial_\mu \chi \partial_\nu\chi - \partial_\chi \W \, F_{\mu\nu}\right) \right)  = 0  \\
& \partial_\mu \left( \frac{\tens}{2\kappa_5^2} \, \Z \,\W \sqrt{-\Gamma} \, \Gamma^{[\mu\nu]} \right)  = 0 \ .
\end{aligned}
\ee
Plugging in the fluctuations given in equation~\eqref{eq.FluctuationAnsatz} we obtain at linear order the following fluctuation equations
\be\label{eq.EquationForFluctuations}
\begin{aligned}
0 = & \partial_r \left( \frac{\sqrt{g_{xx}}}{\sqrt{-g_{tt} \,g_{rr}}} h_{tx}' \right)  + \frac{2\kappa_5^2 \, \charge}{g_{xx}} a_x' \\
&\quad + \left(  \frac{\sqrt{g_{xx}}}{3\sqrt{-g_{tt}g_{rr}}} \phi'^2 - \frac{g_{xx}' {}^2}{\sqrt{-g_{tt} \, g_{xx}^3 \,g_{rr}}} +  \frac{\sqrt{(2 \kappa_5^2 \,\charge)^2 + \left( g_{xx} \right)^3 \tens^2 \W^2 \Z^2}}{\sqrt{g_{rr} + \k \chi'^2}} \frac{\k \, \chi'^2}{3\sqrt{-g_{tt}} g_{xx} \W} \right) h_{tx}  \\
0 = & \partial_r \left( \frac{\W}{2\kappa_5^2 } \frac{\sqrt{-g_{tt}}}{g_{xx}}  \frac{\sqrt{(2 \kappa_5^2 \,\charge)^2 + \left( g_{xx} \right)^3 \tens^2 \W^2 \Z^2}}{\sqrt{g_{rr} + \k \chi'^2}} a_x'  + \frac{\charge}{g_{xx}} h_{tx} \right)  \\
0 =& \charge \left( \E - A_t\, \zeta_x \right) - \frac{\zeta_x}{2\kappa_5^2}\, \frac{g_{xx}^{5/2}}{\sqrt{-g_{tt} g_{rr}}} \partial_r \left( \frac{g_{tt}}{g_{xx}} \right)  \ .
\end{aligned}
\ee

\section{Computations in the D3-D7 system}\label{app:D3D7 computations}

In this second Appendix, we provide details relevant for the D3-D7 system, described in Sec.~\ref{sec:setupD3D7}.

\subsection{Numerical embeddings}

In this Appendix we review the relevant computations we have to do in the D3-D7 probe system in order to find how the D7-branes are embedded in the D3-brane background, which is needed to compute the transport coefficients. We will be following \cite{Kobayashi:2006sb,Mateos:2007vn} closely.

First of all, we need to introduce a new radial coordinate $\zeta$, defined in terms of $r$ through 
\begin{equation}
    r \,=\,  \frac{r_\mt{h}}{\sqrt{2}}\frac{\sqrt{1+\zeta^4}}{\zeta}\,.
\end{equation}
Given that $r_\mt{h}$ is the position of the horizon in the $r$ coordinate, the new coordinate ranges from zero, corresponding to the boundary, to one, where we can find the horizon ($\zeta\in(0,1)$). It is also convenient to work with \footnote{In \cite{Kobayashi:2006sb,Mateos:2007vn} $\chi$ was used for the sine of the angle rather than $\chiang$. Here we changed the notation because $\chi$ is the tachyon in our case, which equals the angle in the D3-D7 system.
}
\begin{equation}
    \chiang(\zeta) \, = \, \sin\left(\theta(\zeta)\right)
\end{equation}
rather than just $\theta$. Using this definition, the first thing we need to find is the profile of $\chiang$, by solving its equation of motion \eqref{eq:D3D7_embedding_eq} in the new variables. Defining
\begin{equation}
    m = \frac{\mtilde \sqrt{2}}{L^2\pi T}\,,\qquad 
c = \frac{\left(6 \ctilde-\mtilde^3\right)\sqrt{2} }{3 \pi ^3 L^6  T^3}\,,
\end{equation}
in terms of $\tilde{m}$ and $\tilde{c}$ from \eqref{eq:D3D7_UV_exp}, the corresponding boundary conditions will be
\begin{equation}
\label{eq:mateos_mass_and_condensate}
\begin{aligned}
    \chiang(\zeta)\, &= \,  m\, \zeta\ + \ c\, \zeta^3\, + \,  \OO(\zeta^4) \\[2mm]
    \chiang(\zeta)\, &= \, \chiang_\mt{h} +\OO(\zeta-1)^1\,,
    \end{aligned}
\end{equation}
near the boundary and the horizon respectively. Because \eqref{eq:D3D7_embedding_eq} only depends on $\chiang(\zeta$) once the equation for the derivative of $A_t$ \eqref{eq:D3D7_electric_field_eq} is assumed, we can easily find solutions. For this, we choose a value of $\chiang_\mt{h}$ at the horizon and solve the equation numerically up to the boundary, where we read off the parameters $m$ and $c$, related to the mass and quark condensate via \eqref{eq:mateos_mass_and_condensate} and \eqref{eq:D3D7_quark_mass_condensate}.

Once the profile of $\chiang$ is known, it is easy to obtain the rest of the quantities that we need. Concerning the chemical potential, note that $A_t$ must vanish at the horizon, so we can find it by direct integration of \eqref{eq:D3D7_electric_field_eq} 
\begin{equation}
    \mu = \frac{r_\mt{h}}{2\pi\ls^2} \, \times 2\,  \tilde{d}\int_{0}^1 \dd \zeta\, \frac{ \, \zeta  \left(1-\zeta ^4\right) \sqrt{1\, -\, \chiang
   ^2\, +\, \zeta ^2 (\chiang')^2}}{\sqrt{\left(\zeta ^4+1\right) \left(1\, -\, \chiang^2\right) \left(8
   \,\tilde{d}^2\, \zeta ^6\,-\,\left(\zeta ^4+1\right)^3 \left(\chiang ^2-1\right)^3\right)}} \ ,
\end{equation}
where the new parameter $\tilde d$ is related to the density
\begin{equation}
    \rho \, = \frac{1}{8}\, \tilde{d}\, \Nc\, \Nf\, T^3\, \sqrt{\lambda} \ .
\end{equation}

Following \cite{Kobayashi:2006sb,Mateos:2007vn}, the entropy density corresponding to the addition of unbackreacted flavor is given by
\begin{equation}\label{eq:sf}
    \, s_\mt{f} \, = \frac{\Nc\Nf\lambda T^3}{64} \,\left(\  -4\tilde{G}(m) \ +\ 12 \tilde{d} \tilde{\mu} \ +\  (m^2 -1)^2\ -\ 6mc\ \right)\,,
\end{equation} 
where 
\begin{equation}\begin{aligned}
    &\tilde{G}(m)=\int_0^1 \dd \zeta\, \left[\frac{ m^2}{\zeta^3}-\frac{1}{\zeta^5}+\frac{\left(1-\zeta ^8\right) \left(1-\chiang ^2\right)}{\zeta ^5}
    \, 
    \sqrt{1+\frac{8\, \tilde{d}^2\, \zeta ^6}{\left(\zeta ^4+1\right)^3 \left(1-\chiang^2\right)^3}} \sqrt{1-\chiang^2+\zeta ^2 (\chiang ')^2}\right]\\[2mm]
\end{aligned}
\end{equation}
and
\begin{equation}
    \tilde{\mu} =  \frac{2\pi\ls^2} {r_\mt{h}} \,\mu= \frac{2}{T\sqrt{\lambda}}\, \mu \ .
\end{equation}

For the computation of the bulk viscosity, we will be interested in the variation of $\chiang_\mt{h}$ with respect to the entropy and the charge densities. For the first one, $\partial\chi_\mt{h} / \partial s$, we consider the variation of the value of the embedding function as we vary the entropy of the background, since including the flavor contribution would be a higher order correction in $\Nf$. On top of that, the entropy density of the background is given in terms of 
\begin{equation}
    s_\mt{g} = \frac{\pi^2}{2}\, \Nc^2\, T^3 \ \  , \ \ T = \frac{2 \sqrt{2} M_q}{\sqrt{\lambda } m} \ ,
\end{equation}
and therefore
\begin{equation}
    s \frac{\partial\chi_\mt{h}}{\partial s}  = \frac{1}{(1-\chiang^2)^{\frac{1}{2}}}
    \left(s \frac{\partial\chiang_\mt{h}}{\partial s}\right) = \frac{1}{(1-\chiang^2)^{\frac{1}{2}}}\left(\frac{m^{-1}}{3}\frac{\partial \chiang_\mt{h}}{\partial(m^{-1})}\right)\,,
\end{equation}
where one has to make sure that $\rho$ is kept fixed as $\chiang_\mt{h}$ and $m^{-1}$ are varied. Similarly, since the other derivative involved in the computation of the bulk viscosity is performed at finite temperature, it can be easily written in terms of $\tilde{d}$ as
\begin{equation}
    \rho \frac{\partial\chi_\mt{h}}{\partial \rho} = \frac{1}{(1-\chiang^2)^{\frac{1}{2}}}
    \left(\tilde{d}\, \frac{\partial\chiang_\mt{h}}{\partial \tilde{d}}\right)\,.
\end{equation}
In conclusion, the equation for the bulk viscosity reduces to
\begin{equation}
    \frac{\zeta}{\eta} \, = \,  \frac{2 \lambda\,}{\pi\Nc^2}\frac{\sigma}{T}\left(
    \frac{m^{-1}}{3}\frac{\partial \chiang_\mt{h}}{\partial(m^{-1})} + 
    \tilde{d} \, \frac{\partial \chiang_\mt{h}}{\partial \tilde{d}}\right)^2\frac{1}{1-\chiang_\mt{h}^2} \ .
\end{equation}

\subsection{Low-temperature computations}

Next, we wish to discuss some technicalities of the zero temperature limit discussed in Sec.~\ref{sec:resultsD3D7}. Recall that we were able to write the entropy as 
\be
 s   =   -\frac{\partial f(T,\rho)}{\partial T}  \ =\  - \pi L^2 \frac{\partial f(r_\mt{h},\rho)}{\partial r_\mt{h}} 
 =   \pi L^2 \tilde{\LL}\, \Big|_{r=r_\mt{h}} -
\pi L^2
\int_{r_h}^\infty \dd r\, \left(\frac{\dd}{\dd r_\mt{h}}\right)_{\rho,m}\left[ \tilde{\LL} (\theta,{\theta'},A_t,A'_t;r_\mt{h}) \right]\ ,  \label{eq_app:entropylowT}
\ee
where the derivative inside the integral is defined through the full variation of the Lagrangian, and we are also keeping the quark mass fixed (even though this is not written down explicitly).

As we shall see next, the second term in the last expression in \eqref{eq_app:entropylowT} vanishes in the limit $r_\mt{h}\to 0$. 
This follows from a standard argument: as we are varying the action around its on-shell value, the variation boils down to boundary terms, which vanish as we are keeping the sources fixed while varying. We will check this now explicitly.
For that, note that at any finite $r_\mt{h}$:
\begin{eqnarray}
\tilde{\LL} (\theta,{\theta'},A_t,A'_t; r_\mt{h}) &=& \cN r^3 \cos^3\theta \left[1-(4\pi\alpha')^2\, (A_t')^2 + r^2 \, f\, (\theta')^2\right]^{\frac{1}{2}}+\rho\,  A_t'\qquad \nonumber\\[2mm] &=& \frac{\cN}{r}\left[\left(  \frac{\rho^2}{(2\pi\alpha')^2\, \cN^2} + r^6 \cos^6\theta\right)(r^2 + (r^4-r_\mt{h}^4)\, \theta'(r)^2)\right]^{\frac{1}{2}}\,.
\end{eqnarray}
Consequently, the derivative of $\tilde\LL$ with respect to $r_\mt{h}$ will give raise to three different terms,
\begin{equation}\label{eq:derrh1}
\int_{r_\mt{h}}^\infty \dd r\, \left[\frac{ \ \partial\tilde{\LL}}{\partial r_\mt{h}} \ +\  
\frac{\partial \tilde{\LL}}{\partial \theta}\,  \partial_{r_\mt{h}}\theta \ + \ 
\frac{ \partial\tilde{\LL}}{\partial\theta'} \, \partial_{r_\mt{h}}\theta' \,
\right]\ .
\end{equation}
In our notation, the first term in this expression represents the derivative of $\tilde{\LL}$ with respect to the explicit dependence on $r_\mt{h}$. Note that it is of order $\OO(r_\mt{h}^3)$ and that it vanishes when $r_\mt{h}$ is taken to zero. Integrating by parts, 
equation \eqref{eq:derrh1} reduces to
\begin{equation}\label{eq:derrh2}
\int_{r_h}^\infty\, \left[ \ \left( \frac{ \partial \tilde{\LL}}{\partial \theta} -\partial_r\,\frac{ \partial \tilde{\LL}}{\partial \theta'} \right)\partial_{r_\mt{h}}\theta \right] \dd r \ + \ \frac{\partial \tilde{\LL} }{\partial\theta' } \, \partial_{r_\mt{h}}\theta \Bigg|_{r_\mt{h}}^\infty\,.
\end{equation}
Here, the integral vanishes on shell. In order to see that the second term vanishes, we need to examine both limits separately. On the one hand, recall that the UV behavior of the embedding function $\theta$ was given in \eqref{eq.assymD3D7}. There, the parameter $\mtilde$ is related to the mass of the quark and will not change as we lower the temperature. In contrast, $\ctilde$ depends on the quark condensate, which varies as we change $r_\mt{h}$. Fortunately, 
\begin{equation}
\partial_{r_\mt{h}} \theta = \frac{\partial_{r_\mt{h}}\ctilde}{r^3}+\ldots\,,
\end{equation}
vanishes when it is evaluated at the boundary. In the opposite limit, it is important to recall that $\theta(r=r_\mt{h})\equiv\theta_\mt{h}$ keeps finite when the limit $r_\mt{h}\to 0$ is taken. This renders $\partial_{r_\mt{h}} \theta $ finite at the horizon. However, we can find a series expansion of the embedding about $r=r_\mt{h}$ of the form
\begin{equation}
\theta = \theta_\mt{h} + \theta_1 (r-r_\mt{h}) \, + \, \ldots
\end{equation}
with
\begin{equation}
    \theta_1\ =\  -\, \frac{3}{4}\, \cdot\, \frac{ (2 \pi \alpha')^2  \, \mathcal{N}^2 \, r_\mt{h}^5 \, \sin \theta_\mt{h}\,  \cos ^5\theta_\mt{h}}{(2\pi \alpha')^2
   \mathcal{N}^2 r_\mt{h}^6 \cos ^6\theta_\mt{h}+\rho ^2}
\end{equation}
which renders ${\partial \tilde{\LL} }/{\partial\theta' }$ linear in $(r-r_\mt{h})$ as
\begin{equation}
    \frac{\partial \tilde{\LL} }{\partial\theta' } \ = \ \frac{6 \pi \alpha '\,  \mathcal{N}^2 \,  r_\mt{h}^6 \, \sin \theta_\mt{h} \, \cos ^5\theta
   _\mt{h}}{\sqrt{\rho ^2\, +\, (2 \pi \alpha')^2 \, \mathcal{N}^2 \,  r_\mt{h}^6 \, \cos ^6\theta _\mt{h}}} \cdot \left(r-r_\mt{h}\right) \ + \ \OO((r-r_\mt{h})^1
\end{equation}
and therefore it is zero when evaluated at the horizon. Put differently, the second term in \eqref{eq:derrh2} is also zero at $r=r_\mt{h}$, even before taking the $r_\mt{h}\to 0$ limit. 

Knowing that the first tem in \eqref{eq_app:entropylowT} cancels, the contribution to entropy density from the flavor sector at zero temperature can be found. The computation is finished in Sec.~\ref{sec:resultsD3D7}. Moreover, with the zero temperature embeddings given there it is possible to evaluate $S_{\text{DBI}}$, which gives \cite{Karch:2007br}
\be
\frac{S_{\text{DBI}}}{\beta V_3} = - \Omega = p = \frac{\Nc\, \Nf\, \gamma}{4\lambda} (\mu^2-\mass^2)^2\,, 
\ee
from which the energy density can be also obtained
\be
\varepsilon \, =\,  \mu\rho - p \, =\,  \frac{\Nf\, \Nc\, \gamma}{4\lambda}\, (3\mu^2+\mass^2)(\mu^2-\mass^2)\,,
\ee
where we have also used \eqref{eq:D3D7density_and_C_low_temp}. From this, the stiffness (speed of first sound squared) reads \cite{Kulaxizi:2008jx,Kim:2008bv}:
\be \label{eq_app:speedofsoundD3D7}
c_s^2 = \frac{\partial p}{\partial \varepsilon}\, \bigg|_{s} = \frac{\partial p}{\partial \mu}\cdot \left(\frac{\partial \varepsilon}{\partial \mu}\right)^{-1}\, =\, \frac{\mu^2-\mass^2}{3\mu^2-\mass^2}\ .
\ee

\section{Definitions for the V-QCD model}\label{app:vqcd}

In this Appendix, we write down the detailed definitions for the V-QCD model. We start from the action for the glue sector, {\emph{i.e.}}, the action for the dilaton field $\lambda = \exp(\sqrt{3}\phi/\sqrt{8})$. The main input is the dilaton potential, which we choose to be (following~\cite{Alho:2015zua,Jokela:2018ers})
\begin{equation}
    V(\lambda) = -12\left[1 + V_1 \lambda + V_2 \frac{\lambda^{2}}{1+\lambda/\lambda_0} +V_\mathrm{IR} e^{-\lambda_0/\lambda}(\lambda/\lambda_0)^{4/3}\sqrt{\log(1+\lambda/\lambda_0)}\right] \ .
\end{equation}
Notice that this definition satisfies the asymptotic requirements in~\eqref{eq:VUVlim}
and in~\eqref{eq:VIRlim}. 
The parameters $V_1$ and $V_2$ are determined by matching the holographic RG flow of the coupling with  the running of the coupling in pure Yang-Mills theory~\cite{Gursoy:2007cb,Jarvinen:2011qe}. That is, the model implements asymptotic freedom -- it is in principle put in by hand, but arises for a natural choice of potentials, {\emph{i.e.}}, potentials which are analytic in $\la$ at the UV point $\la=0$. The matching leads to
\begin{equation}
    V_1 = \frac{11}{27\pi^2} \ , \qquad V_2 = \frac{4619}{46656 \pi ^4} \ ,
\end{equation}
where we dropped terms suppressed by $1/\Nc^2$. 
We stress that the physically more relevant part is the behavior of the potential at large $\la$, and the asymptotics was chosen to reproduce several known IR features of pure Yang-Mills theory: confinement, magnetic screening, and linear radial Regge trajectories for the glueballs. 
Finally, the remaining parameters  $\lambda_0$ and $V_\text{IR}$ are determined by comparing the behavior of the model at intermediate energies to lattice data. Specifically, we used the data for the thermodynamics of the Yang-Mills theory at large $\Nc$ from~\cite{Panero:2009tv}, following the approach of~\cite{Gursoy:2009jd}. This fit leads to~\cite{Alho:2015zua,Jokela:2018ers,Alho:2020gwl}
\begin{equation}
  \lambda_0 = 8\pi^2/3 \ , \qquad V_\mathrm{IR} = 2.05 \ .
\end{equation}

We then discuss the potentials of the flavor sector, {\emph{i.e.}}, the potentials $\Z$, $\W$, and $\kappa$  in Eq.~\eqref{eq:ActionDefsFlavor}. Writing $\Z(\lambda,\chi)= V_{{\mt{f}}0}(\lambda)e^{-\chi^2}$, we use the following Ansatz:
\begin{align} \label{eq:Vf0fit}
     V_{{\mt{f}}0}(\lambda) &= W_0 + W_1 \lambda +\frac{W_2 \lambda^2}{1+\lambda/\lambda_0} + W_\mathrm{IR} e^{-\lambda_0/\lambda}(\lambda/\lambda_0)^{2}  & \\
\frac{1}{\kappa(\lambda)} &= \kappa_0 \left[1+ \kappa_1 \lambda + \bar \kappa_0 \left(1+\frac{\bar \kappa_1 \lambda_0}{\lambda} \right) e^{-\lambda_0/\lambda }\frac{(\lambda/\lambda_0)^{4/3}}{\sqrt{\log(1+\lambda/\lambda_0)}}\right]  & \\
\frac{1}{\W(\lambda)} &=  w_0\left[1 + \frac{w_1 \lambda/\lambda_0}{1+\lambda/\lambda_0} + 
\bar w_0 
e^{-\hat\lambda_0/\lambda}\frac{(\lambda/\hat\lambda_0)^{4/3}}{\log(1+\lambda/\hat\lambda_0)}\right] \  .
\label{eq:wfit}
\end{align}
Notice that we do not need the potential $\kappa$ in this article, because for V-QCD we only consider chirally symmetric background solutions with zero quark mass so that $\chi=0$ and consequently the results are independent of $\kappa$. We however discuss a consistent choice of $\kappa$ here for completeness.
As we explained in the main text, the asymptotics of the potentials were determined by several requirements. A ``good'' kind of IR singularity in the classification of~\cite{Gubser:2000nd}, vanishing of the flavor action at the bottom for the chirally broken solutions~\cite{Jarvinen:2011qe}, consistent solutions even at finite $\theta$-angle~\cite{Arean:2016hcs}, and linear radial meson trajectories~\cite{Arean:2013tja} were found when $V_{{\mt{f}}0}(\la) \sim \la^{v_p}$ with $v_p \approx 2$ and $\kappa(\la) \sim \la^{-4/3}\sqrt{\log \la}$. In order for the vector and axial mesons to have the same slope $\kappa(\la)$ needs to vanish faster than $\W(\la)$ at large $\la$~\cite{Arean:2013tja}, and requiring for a consistent phase diagram at finite density fixes the power law in the asymptotics of $\W$ to be the same as in the asymptotics of $\kappa$, {\emph{i.e.}}, $\W(\la) \sim \la^{-4/3}$~\cite{Ishii:2019gta}. 

The parameters of the fits which control the expansion at small $\la$ were determined by requiring agreement with the UV dimension of the $\bar qq $ operator and the perturbative two-loop running of the 't Hooft coupling in QCD in the Veneziano limit (see the discussion at the end of this Appendix). Setting $x_\mt{f}=\Nf/\Nc$ to one, this gives
\begin{equation}
 \kappa_0 = \frac{3}{2} - \frac{W_0}{8} \ , \qquad W_1 = \frac{8+3\, W_0}{9 \pi ^2} \ , \qquad W_2 = \frac{6488+999\, W_0}{15552 \pi ^4} \ .
\end{equation}

\begin{table}
\caption{Fit to thermodynamics at small chemical potential: values of various parameters. Here $L$ is the UV AdS radius and $M^3$ was normalized such that the tabulated value gives the deviation from the Stefan-Boltzmann law for the pressure at high temperatures. } \label{tab:thermofit}
\begin{center}
\begin{tabular}{|c||c|c|c|}
\hline
 & soft & intermediate  & stiff \\[-1mm]
 & \textbf{5b} & \textbf{7a}  & \textbf{8b} \\
\hline
\hline
$W_0$ & 1.0 & 2.5 &  5.886 \\
\hline 
\hline 
$W_\mathrm{IR}$ & 0.85 & 0.9 & 1.0 \\
\hline
$w_0$ & 0.57 & 1.28 & 1.09 \\
\hline
$w_1$ & 3.0 & 0 & 1.0 \\
\hline
$\bar w_0$ & 65 & 18 & 22 \\
\hline
$8 \pi^2/\hat \lambda_0$ & 0.94 & 1.18 & 1.16 \\
\hline
\hline 
$\bar\kappa_0$ & 1.8 & 1.8 & 3.029 \\
\hline
$\bar\kappa_1$ & -0.857 & -0.23 & 0 \\
\hline
\hline 
$\Lambda_\mathrm{UV}$/MeV & 226 & 211 & 157 \\
\hline
$180 \pi^2 M^3L^3/11$ & 1.34 & 1.32 & 1.22 \\
\hline

\end{tabular}
\end{center}
\end{table}

The remaining parameters in~\eqref{eq:Vf0fit} and~\eqref{eq:wfit} were determined by comparing to the lattice data of full QCD with 2+1 flavors from~\cite{Borsanyi:2013bia,Borsanyi:2011sw}. Specifically, the parameters of the $V_{{\mt{f}}0}$ potential (as well as the Planck mass $M_\text{Pl}$) to the data for the interaction measure $(\eps-3p)/T^4$ and the parameters of the $\W$ potential to the data for the first cumulant $\chi_2=\partial^2_\mu p(T,\mu)|_{\mu=0}$ of the pressure, {\emph{i.e.}}, the baryon number susceptibility. The parameters of the $\kappa$ potential were chosen such that the temperature of the phase transition is consistent with the fit to the interaction measure. See~\cite{Jokela:2018ers} for details. As the fit has a flat directions, it does not fully constrain the parameter values leaving free what is essentially a one-parameter family of the solutions. We have taken this remaining parameter as $W_0$, and chosen three fits (soft, intermediate, and stiff variants) corresponding to different values of $W_0$. The fit parameters are given in Table~\ref{tab:thermofit}. The names refer to the stiffness of the equation of state for the nuclear matter phase (that we do not discuss in this article, see~\cite{Ishii:2019gta,Ecker:2019xrw,Jokela:2020piw,Demircik:2020jkc}), with stiff (soft) equation of state meaning that the speed of sound is high (low), and the stiffness increases with $W_0$ \footnote{The equations of state at zero temperature have already been published in the CompOSE online repository \href{https://compose.obspm.fr/}{https://compose.obspm.fr/}~\cite{Typel:2013rza}. The finite temperature equations of state, along with the transport quantities, are being published there as the preprint version of this work is being finalized.}. The soft, intermediate, and stiff variants were labeled as the fits  \textbf{5b}, \textbf{7a}, and \textbf{8b}, respectively in~\cite{Jokela:2018ers} (the values of the $\kappa$ function were slightly adjusted in~\cite{Ishii:2019gta,Jokela:2020piw} to make sure that the chirally broken solutions are consistent in the IR). 

There is one more parameter which needs to be determined, namely the overall energy scale of the theory. As in QCD, the energy scale does not appear as a parameter in the action but is a property of the solutions. We choose to determine it by using the scale of the UV expansions of the background solutions $\Lambda_\mathrm{UV}$, which corresponds to $\Lambda_{QCD}$ in QCD, but differs by an unspecified numerical constant. The QCD scale and running coupling have also been discussed in a different holographic framework in~\cite{Brodsky:2010ur}.

In order to define the weak coupling scale $\Lambda_\mathrm{UV}$ we employ the gauge $g_{tt}(r)g_{rr}(r)=g_{xx}(r)^2$. In this gauge, we find that
 \begin{align} \label{eq:UVmetric}
    g_{xx}(r) &= \frac{L^2}{r^2}\left[1+\frac{8}{9 \log r \Lambda_\text{UV}} +\mathcal{O}\left(\frac{1}{(\log r\Lambda_\text{UV})^2}\right)\right]  & \\
    \lambda(r) &= e^{\sqrt{3/8}\, \phi(r)} = -\frac{8\pi^2}{3 \log r \Lambda_\text{UV}}   -\frac{28 \pi ^2}{27}  \frac{\log(-\log r\Lambda_\text{UV} )}{(\log r \Lambda_\text{UV})^2} +   \mathcal{O}\left(\frac{1}{(\log r\Lambda_\text{UV})^3}\right) \ , \label{eq:UVlambda}
\end{align}
where $L = 1/\sqrt{1-W_0/12}$ is the asymptotic AdS radius, and the boundary is reached as $r \to 0$ from above. In order to match these expansions with the RG flow of QCD, we note that the energy scale is dual to $\sqrt{g_{xx}(r)}$~\cite{Gursoy:2007cb}. Then it is straightforward to check that the RG flow of the coupling indeed matches with the of QCD (which was ensured by the choice of the parameters $V_{1,2}$ and $W_{1,2}$ above). Finally, comparing to the lattice fit of the interaction measure, we find the values of $\Lambda_\mathrm{UV}$ given in Table~\ref{tab:thermofit}.

\section{V-QCD near the zero temperature AdS$_2$ point} \label{app:ads2}

In this fourth Appendix, we specialize to the low-temperature behavior of the V-QCD model.

\subsection{Analysis of the backgrounds}

We discuss here only the case of vanishing tachyon, $\chi=0$. Writing the metric as
\be \label{eq:metricAn}
 ds^2 = b^{2}\left(dr^2/f-f dt^2+d\mathbf{x}^2\right)= e^{2A}\left(dr^2/f-f dt^2+d\mathbf{x}^2\right) \,,
\ee
the gauge field equation of motion can be integrated:
\be
{-
bV_\mt{f}\W^2 \Azs'\over
\sqrt{1
-{\W^2\over b^4}
(\Azs')^2}}=\hat n \,,
\ee
where $\hat n$ is a constant related to the charge density through
\be
 \rho = \frac{1}{2\kappa_5^2} \hat n \ . 
\ee
It turns out to often be convenient to use instead the dimensionless quantity 
\be 
\nt = e^{-3 A_\mt{h}} \hn = \frac{\hn}{b_\mt{h}^3} = \frac{4 \pi \rho}{s} \ ,
\ee  
where the subscript ``$\textrm{h}$'' refers to the value at the horizon. 

After eliminating the gauge field, the equations of motion can be written in a simple form in terms of the effective potential 
\be\label{eq:Veff}
 V_\mathrm{eff}(\l,A,\hn) = -V(\l)-\tens V_{{\mt{f}}0}(\l)\sqrt{1+\frac{\hn^2}{e^{6A} \W(\l)^2 x^2 V_{{\mt{f}}0}(\l)^2}}\ ,
\ee
namely
\bea \label{feqtau0}
f''+3 A'   f' &=& \frac{1}{3} e^{2 A} \frac{\pa V_\mathrm{eff}}{\pa A} \\ 
A''-\left(A'\right)^2+\frac{4 \left(\lambda '\right)^2}{9
   \lambda ^2}&=&0 \\
 \frac{3 A' f'}{f}+12  
\left(A'\right)^2-\frac{4 \left(\lambda '\right)^2}{3 \lambda
   ^2}&=&\frac{e^{2 A} V_\mathrm{eff}}{f} \ .
   \label{eq:ceq}
\eea
The equation for the dilaton is not independent: 
\be
\label{laeqtau0}
\lambda
   ''+3 A' \lambda '+\frac{f' \lambda '}{f}-\frac{\left(\lambda '\right)^2}{\lambda }=-\frac{3 e^{2 A} \lambda ^2}{8 f}\frac{\pa V_\mathrm{eff}}{\pa \l} \,.
   \ee
The equations of motions are unchanged under shifts of the coordinate $r$, but there is also another symmetry which takes $f \mapsto f_0 f$ and $A \mapsto A + (\log f_0)/2$.

The tachyonless black hole solutions are conveniently indexed in terms of the horizon parameters $\lambda_\mt{h}$ and $\nt$. After numerically constructing the full solution, these parameters can be mapped to the temperature and the chemical potential~\cite{Alho:2013hsa}.

\begin{figure}[!t]
\begin{center}
\includegraphics[width=0.9\textwidth]{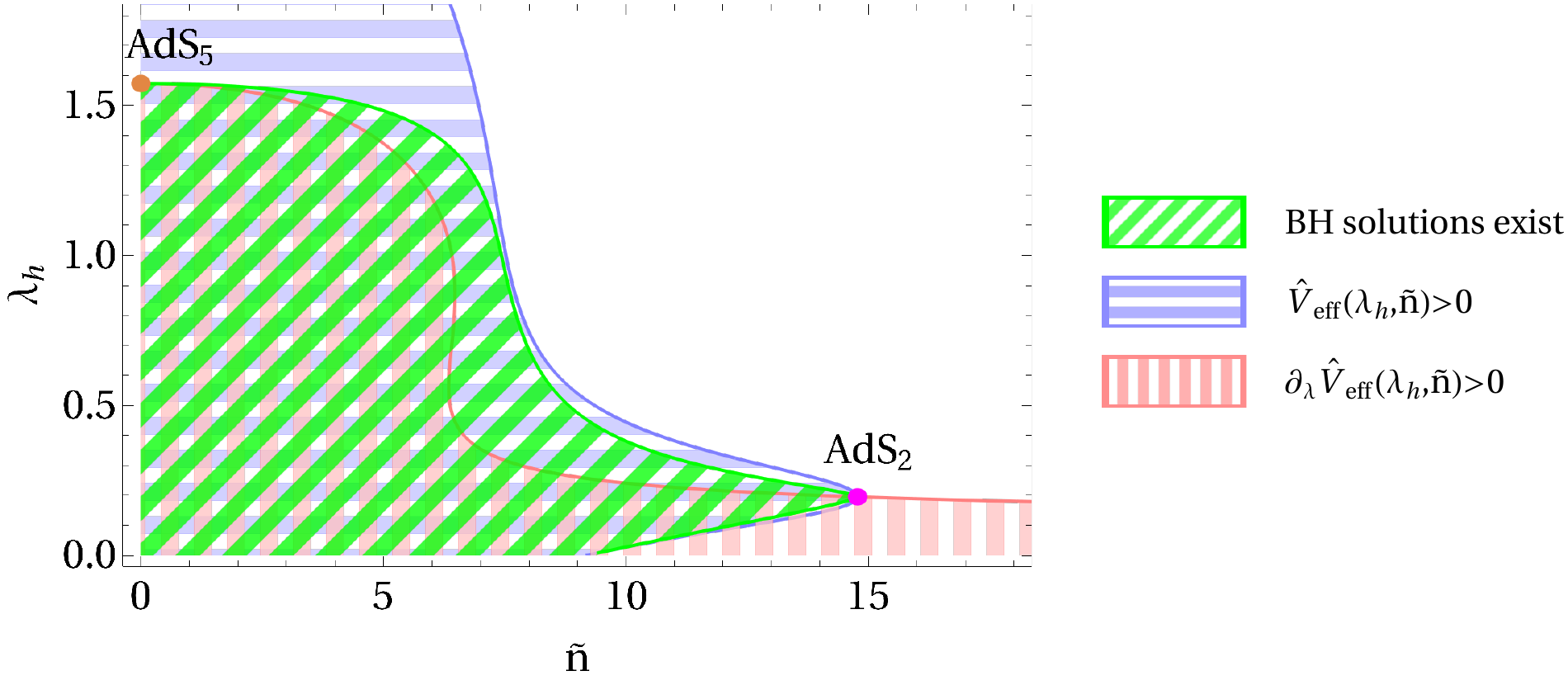}
\end{center}
\caption{Behavior of the effective potential, and the location of fixed points for potentials {\bf{7a}}. Blue, and red striped regions have $\hat V_\mathrm{eff}>0$ and $\partial_\la \hat V_\mathrm{eff}>0$, respectively, while the green region marks where the chirally symmetric BH solutions exist. The left (orange) and right (magenta) dots mark the (unstable) AdS$_5$ and (stable) AdS$_2$ fixed points, respectively.  }
\label{fig.7aVeff} 
\end{figure}

The study the fixed point solutions to the equations of motion~\eqref{feqtau0}--\eqref{eq:ceq} boils down to the properties of the effective potential $V_\mathrm{eff}$ at the horizon $A=A_\mt{h}$ (see~\cite{Alho:2013hsa}). Notice first that in terms of  $\tilde n = \hn e^{-3A_\mt{h}}$, the effective potential
\be 
V_\mathrm{eff}(\la ,A_\mt{h},\tilde n) = -V(\l)-\tens V_{{\mt{f}}0}(\l)\sqrt{1+\frac{\nt^2}{\W(\l)^2 x^2 V_{{\mt{f}}0}(\l)^2}} \equiv \hat V_\mathrm{eff}(\la,\nt)
\ee
is independent of $A_\mt{h}$.
The AdS$_2$ solution is found at the fixed points satisfying
\be
 \hat V_\mathrm{eff}(\la_*,\tilde n_*) = 0 =\pa_\la \hat V_\mathrm{eff}(\la_*,\tilde n_*) \,.
\ee
These equations typically have one or zero solutions $(\la_*,\nt_*)$ for physically reasonable potentials, but may also have two (or perhaps even more) solutions. In Fig.~\ref{fig.7aVeff}, we show the single solution for potentials {\bf{7a}} as the magenta dot. It is found at the intersection of the blue and red curves, which mark where $\hat V_\mathrm{eff}$ and $\partial_\la \hat V_\mathrm{eff}$ vanish, respectively. Chirally symmetric BH solutions are only found in a region of the $(\nt,\lambda_\mt{h})$ -plane around the origin~\cite{Alho:2013hsa} (region with green stripes in Fig.~\ref{fig.7aVeff}). The AdS$_2$ fixed point(s) lie on the boundary of this region. Therefore all the curves in the figure pass through this point. Notice that $\lambda_\mt{h} > 0$ and we may take $\nt \ge 0$ by symmetry. 

In general, the zero temperature solutions are found at the edge of the region where regular BH solutions exist (green curve in the figure). They are given as holographic RG flows from the AdS$_5$ geometry in the boundary (given by Eqs.~\eqref{eq:UVmetric} and~\eqref{eq:UVlambda}) to an IR geometry which depends on the location on the curve. These zero temperature solutions may be classified as follows:
\begin{itemize}
 \item Zero temperature solutions at the AdS$_2$ points. Even though these configurations are single points on the $(\nt,\lambda_\mt{h})$ -plane, they are mapped to finite intervals of the chemical potential. The solutions can be obtained through a limiting procedure for the BH solutions (as seen numerically): different values of chemical potentials are found depending on the direction on the $(\nt,\lambda_\mt{h})$-plane from where the fixed point is approached. The geometry is asymptotically AdS$_2$ in the IR, as we will discuss in detail below. Notice that apart from the flow solutions, there is also an exact AdS$_2$ solution with constant dilaton which is not identified with any physical phase on the field theory side.
 \item The zero temperature solution at $\nt =0$ and $\lambda_\mt{h}=\la_*$. This is a chirally symmetric special solution at zero temperature and density, which is subdominant for all potentials which we consider here (as should be the case for QCD since the vacuum is chirally broken). This point corresponds to an AdS$_5$ IR geometry.  
 \item Solutions elsewhere on the edge. We have found numerically that for some choices of the potentials, the AdS$_2$ points do not cover all values of positive chemical potential, but in particular zero temperature solutions for high values of $\mu$ are found elsewhere on the edge of the region where solutions exist. These solutions and the transition where one moves away from AdS$_2$ merits further study which we leave for future work. It is not even completely clear that solutions of this type with strictly $T=0$  exist -- our studies are currently  restricted to numerical BH solutions with tiny but nonzero temperature, and taking the $T \to 0$ limit numerically is tricky. 
\end{itemize}

\subsection{Flow to the fixed point solution}

Fluctuations around the exact AdS$_2$ fixed point geometry have been studied in~\cite{Alho:2013hsa}. Two normalizable modes were found, one having the dimension one and another with a nontrivial dimensions denoted by $\alpha_-^\la$ in section 3 of this article. In our notation we may define
\be
\alpha = \frac{1}{2}\left[-1+\sqrt{1-\frac{9 \la_*^2 \pa^2_\la V_\mathrm{eff}(\la_*,A_\mt{h},\nt_*)}{\pa_A V_\mathrm{eff}(\la_*,A_\mt{h},\nt_*)}}\right]\,.
\ee
Notice that here the dependence on $A_\mt{h}$ cancels after $\hn$ is replaced by $\nt$. 

The solution around the fixed point, i.e., the holographic RG flows ending in the AdS$_2$ geometry, may then be found in terms of transseries
\be
 g(r) = \sum_{i,j = 0}^{\infty} g_{ij} C^{j} \left((r_0-r)\Lambda_\mathrm{IR}\right)^i \left((r_0-r)\Lambda_\mathrm{IR}\right)^{\alpha j} 
\ee
for the fields $g = \la,\ f$, and $A$. Here $r_0$ is the value of the coordinate where the fixed point is reached, and $\Lambda_\mathrm{IR}$ and $C$ are constants which appear due to symmetries of the equations of motion. The constant $\Lambda_\mathrm{IR}$ is interpreted as the energy scale and $C$ will be mapped to the value of the chemical potential.

The coefficients $g_{ij}$ may be solved iteratively by inserting the transseries into the equations of motion~\eqref{feqtau0}--\eqref{eq:ceq}. First few terms in the series are 
\begin{align}
\label{eq:laads2}
  \la(r) &=  \la_* + C \hat r^\alpha  +  \left[-\frac{9 \la_*^2 \pa_\la^3 V}{8 \alpha \pa_A V}+\frac{3 \left(-5 \alpha ^2+\alpha +2\right) \pa_\la \pa_A V}{2 (\alpha +1) (\alpha +2) (2 \alpha -1)  \pa_A V}+\frac{3 \alpha +2}{\la_*} \right]
  \frac{C^2 \hat r^{2\alpha}}{1+3\alpha} & \nonumber\\
  &\quad +\morder{\hat r^{3\alpha}} + \frac{9 \la_*^2  \pa_\la \pa_A V}{4 \left(\alpha +2\right)\left(\alpha -1\right) \pa_A V} \hat r + \morder{\hat r^{1+\alpha}} + \morder{\hat r^2}\\
\label{eq:Aads2}  A(r) &= \log f_0+\log\Lambda_\mathrm{IR}+\frac{2 \alpha }{9(1-2 \alpha ) \la_*^2}C^2 \hat r^{2\alpha} + \morder{\hat r^{3\alpha}} \nonumber\\
  &\quad  + \hat r -\frac{2 \pa_\la \pa_A V}{(\alpha +1) \left(\alpha +2\right)\left(\alpha -1\right) \pa_A V} C \hat r^{1+\alpha} +  \morder{\hat r^{2+\alpha}} & \nonumber \\ 
  &\quad  +\left[\frac{1}{2}-\frac{9 (\la_*)^2\left( \pa_\la \pa_A V\right)^2}{8 \left(\alpha +2\right)^2\left(\alpha -1\right)^2 \left(\pa_A V\right)^2}\right] \hat r^2 + \morder{\hat r^{2+\alpha}} & \\
  f(r) &= f_0^2 \hat r^2 \bigg\{\frac{1}{6} \pa_A V+\frac{ \pa_\la \pa_A V }{3\left(\alpha +2\right) \left(\alpha +1\right)}C \hat r^{\alpha} + \morder{\hat r^{2\alpha}} & \nonumber\\
 &\quad \! +\left[\frac{1}{18} \left(\pa_A^2 V-\pa_A V\right)+\frac{\la_*^2 (\pa_\la \pa_A V)^2}{8 \left(\alpha +2\right)\left(\alpha -1\right) \pa_A V}\right] \hat r + \morder{\hat r^{1+\alpha}}  + \morder{\hat r^2}\!\bigg\} \ .
 \label{eq:fads2}
\end{align}
Here $\hat r = (r_0-r) \Lambda_\mathrm{IR}$, $V=V_\mathrm{eff}$, derivatives of the potential are evaluated at the fixed point (also setting $A=A_\mt{h}$), and $f_0$ is another coefficient related to the aforementioned symmetry of the equations of motion for homogeneous configurations. It will be determined by setting $f \to 1$ at the boundary for the full solution. Notice that we chose the coefficients $\lambda_{01} =1$ and $A_{10}=1$ in order to pin down the definitions of $C$ and $\Lambda_\mathrm{IR}$. Moreover we used the fact that $V=0=\pa_\la V$ at the fixed point. The exact AdS$_2$ solution is obtained from these expansions in the limit $\Lambda_\mathrm{IR} \to 0$ keeping the product $f_0 \Lambda_\mathrm{IR}$ fixed so that the flow is suppressed (the coordinate $r$ should also be kept fixed -- recall that there are also factors of $\Lambda_\mathrm{IR}$ hidden in the definition of $\hat r$). 

The full solutions ending at the AdS$_2$ fixed point in the IR can then be constructed numerically by using the expansions as initial conditions. This way one can obtain solutions in a range of chemical potentials which have vanishing temperature.

\subsection{Thermodynamics of small black holes}\label{sec:thermoVQCD}

The relevant geometry for the V-QCD solutions at small temperature and finite density is (in the case an AdS$_2$ fixed point exists) a ``small'' \footnote{Notice that the black hole will not be small in the sense that the area of the black hole vanishes as $T \to 0$: as we shall see below, it goes to a constant instead. Therefore the black holes are ``small'' only in the sense that they are obtained from the AdS$_2$ flow geometry by adding a horizon very close to the IR end, {\emph{i.e.}}, $\hat r_\mt{h}$ defined below is small.} AdS$_2$ black hole. The geometry of such black holes is obtained from~\eqref{eq:laads2}--\eqref{eq:fads2} by adding the AdS$_2$ thermal factor:
\begin{align}
 \la(r) &= \la_* + \morder{\Lambda_\mathrm{IR}} & \\
 A(r) &= \log (f_0\Lambda_\mathrm{IR})  + \morder{\Lambda_\mathrm{IR}} & \\
 f(r) &= \frac{1}{6} \pa_A V f_0^2 \hat r^2 \left(1 - \frac{\hat r_\mt{h}}{\hat r}\right) + \morder{\Lambda_\mathrm{IR}} \ ,
\end{align}
where we only included the leading terms in the AdS$_2$ limit $\Lambda_\mathrm{IR}\to 0$. Higher order corrections can in principle be computed systematically but the computation is quite tedious and after adding the blackening factor the result is no longer a transseries. 

The Hawking temperature of the black hole is given by
\be
 T = - \frac{f'(r_\mt{h})}{4\pi} = \frac{\Lambda_\mathrm{IR}}{4\pi}\, \frac{df}{d\hat r}\Big|_{\hat r= \hat r_\mt{h}} \approx \frac{1}{24 \pi} \pa_A V f_0^2 \Lambda_\mathrm{IR} \hat r_\mt{h} \equiv C_T \hat r_\mt{h} \ .
\ee
Inserting this in the zero temperature asymptotic formula~\eqref{eq:Aads2}, {\emph{i.e.}}, neglecting the backreaction of the blackening factor, we can estimate the leading corrections to the entropy:
\be
 4 G_5 s = e^{3A(r_\mt{h})} \approx f_0^3\Lambda_\mathrm{IR}^3 \left[1 +\frac{2 \alpha }{3(1-2 \alpha ) \la_*^2}C^2 C_T^{-2\alpha} T^{2\alpha} + 3T/C_T + \morder{{T^{3\alpha},T^{1+\alpha},T^2}} \right] \ .
\ee
Taking the backreaction into account is expected to modify the coefficients but not the powers. Since $\eta/s = 1/(4\pi)$ in our model, the shear viscosity will have the same temperature dependence as the entropy.

Evaluating the low temperature behavior of the bulk viscosity is a bit more involved. Instead of using the Eling-Oz formula~\cite{Eling:2011ms}, which expresses the bulk viscosity in terms of the horizon behavior of the background, we find it easier to study the fluctuations. In our case the relevant fluctuation equation reads (see~\cite{Gubser:2008sz,Buchel:2011wx} and Appendix~\ref{app:finiteom})
\be
 h'' + \left(3 A'+ \frac{f'}{f} + 2\frac{X'}{X} \right)h' +\left(
 - \frac{f'}{f}\frac{X'}{X}-\frac{f''}{f} -\frac{3\la e^{2A}}{8 f X} \partial_\la \partial_A V_\mathrm{eff} \right)h  = 0 \ ,
\ee
where we already set the momentum and frequency of the fluctuations to zero -- this is enough for the computation the bulk viscosity. The field $h$ is the fluctuation of the metric measuring the change in spatial volume, defined through $\delta g_{ij}(r) = e^{2A(r)}\delta_{ij}h(r)$ and $X(r) = \la'(r)/(A'(r)\la(r))$.

The bulk viscosity is then obtained as
\be\label{eq:bulkfluct}
 \frac{\zeta}{s} = \frac{2}{27\pi} \frac{h(r_\mt{h})^2\la'(r_\mt{h})^2}{\la(r_\mt{h})^2A'(r_\mt{h})^2} \ ,
\ee
where $h$ is the IR regular solution normalized to $h \to 1$ at the boundary. 

For the zero temperature flow the leading terms close to the AdS$_2$ point give
\be
 \frac{d^2h}{d\hat r^2} + \frac{2\alpha}{\hat r} \frac{dh}{d\hat r} -\frac{2\alpha}{\hat r^2}\, h \approx 0
\ee
with the solution
\be
 h(\hat r) = C_1 \hat r +C_2 \hat r^{-2\alpha} \ ,
\ee
where the first term is the IR regular term. Going to finite temperature, we therefore expect $h(r_\mt{h})\sim \hat r_\mt{h} \sim T$. Inserting the results for the asymptotic AdS$_2$ geometry in~\eqref{eq:bulkfluct} gives
\begin{align}
    \zeta &\sim T \ ,& \qquad & (0<\alpha< 1/2) \ ; &\\
    \zeta &\sim T^{2\alpha} \ ,& \qquad & (1/2<\alpha< 1) \ . &\\
\end{align}

The temperature dependence of the conductivities is simple: using the above results in the formulas~\eqref{eq.DCconductivity} and~\eqref{eq.kappa_equation}  we find that
\be
  \sigma^{xx}\sim \kappa^{xx} \sim T 
\ee
with corrections suppressed by $T^{\alpha}$ as $T \to 0$.

\section{Fluctuations at finite frequency}\label{app:finiteom}

In this final Appendix, we discuss fluctuations of the system and show how the retarded correlators 
\be 
 G_{\eta}^R(\omega) = -i \int dt d^3\vec{x} e^{i\omega t} \theta(t) \langle T_{xy}(t,\vec x)T_{xy}(0,\vec 0) \rangle
\ee
and 
\be 
 G_{\zeta}^R(\omega) = -\frac{i}{9} \int dt d^3\vec{x} e^{i\omega t} \theta(t) \langle {T_{j}}^j(t,\vec x){T_{k}}^k(0,\vec 0) \rangle \ ,
\ee
which are discussed in Sec.~\ref{sec:resultsVQCD}, are computed (see, {\emph{e.g.}},~\cite{Gubser:2008sz}).

For the shear sector we consider, assuming the metric Ansatz~\eqref{eq:metricAn}, fluctuations of the form 
\be
\delta g_{xy}(r,t) = \delta g_{yx}(r,t)= e^{2A(r)} e^{-i\omega t} h_{xy}(r) \ . 
\ee
They satisfy the following fluctuation equation:
\be
h_{xy}''(r) + 3 A'(r) h_{xy}'(r)+\frac{f'(r) h_{xy}'(r)}{f(r)}+\frac{\omega ^2 h_{xy}(r)}{f(r)^2} = 0 \ .
\ee
Let us construct solutions to this equation with the boundary conditions $h_{xy}(0)=1$ and infalling conditions at the horizon. The retarded correlator is then found to be
\be 
 G_\eta^R(\omega) = - \frac{1}{16\pi G_5} \lim_{r\to 0} e^{3A(r)} h_{xy}'(r) \ . 
\ee

For the bulk sector we consider fluctuations of the form 
\be
\delta g_{ij}(r,t) =  e^{2A(r)} \delta_{ij} e^{-i\omega t} h(r) \ . 
\ee
They turn out to satisfy (assuming chirally symmetric vacuum, $\chi=0$) the equation
\be
 h'' + \left(3 A'+ \frac{f'}{f} + 2\frac{X'}{X} \right)h' +\left(\frac{\omega^2}{f^2}
 - \frac{f'}{f}\frac{X'}{X}-\frac{f''}{f} -\frac{3\la e^{2A}}{8 f X} \partial_\la \partial_A V_\mathrm{eff} \right)h  = 0 \ ,
\ee
where $X(r) = \la'(r)/(A'(r)\la(r))$ and $V_\mathrm{eff}(\l,A,\hn)$ was defined in~\eqref{eq:Veff}. Setting again $h(0)=1$ at the boundary and infalling conditions at the horizon, the relevant correlator is found as
\be 
 G_\zeta^R(\omega) = -\frac{1}{16\pi G_5} \lim_{r\to 0} e^{3A(r)} h'(r) \ . 
\ee

\bibliographystyle{apsrev4-1}

\bibliography{biblio}

\end{document}